\documentclass[preprint,aps,amsmath,superscriptaddress,nofootinbib,tightenlines]{revtex4}
\usepackage{bm}
\usepackage{epsfig}

\newif\ifpdf
\ifx\pdfoutput\undefined
\pdffalse 
\else
\pdfoutput=1 
\pdftrue
\fi

\def\bsigma{\mbox{\boldmath $\sigma$}}

\def\OMIT#1{}

\newcommand{\nn}{\nonumber} 

\newcommand{\bn}{{\bar n}}
\newcommand{\bea}{\begin{eqnarray}}
\newcommand{\eea}{\end{eqnarray}}
\newcommand{\nb}{\bar n}

\newcommand{\bnP}{\bar {\cal P}}
\newcommand{\ppP}{{\cal P}_\perp}

\newcommand{\mpsi}{M_\psi}

\newcommand{\jpsi}{J/\psi}
\newcommand{\emax}{E_{max}}

\begin{document}

\ifpdf
\DeclareGraphicsExtensions{.pdf, .jpg}
\newcommand{\picspace}{\vspace{-2.5in}}
\newcommand{\picspacehalf}{\vspace{-1.75in}}
\else
\DeclareGraphicsExtensions{.eps, .jpg,.ps}
\newcommand{\picspace}{\vspace{0in}}
\newcommand{\picspacehalf}{\vspace{0in}}
\fi


\preprint{\vbox{ \hbox{CMU-HEP-03-06}   \hbox{FERMILAB-Pub-03/069-T} }}

\title{Resumming the color-octet contribution to $e^+ e^- \to \jpsi + X$} 

\author{Sean Fleming\footnote{Electronic address: spf@andrew.cmu.edu}}
\affiliation{Department of Physics, Carnegie Mellon University,
      	Pittsburgh, PA 15213
	\vspace{0.2cm}}
	
\author{Adam K. Leibovich\footnote{Electronic address: adam@fnal.gov}}
\affiliation{Department of Physics and Astronomy, 
	University of Pittsburgh,
        Pittsburgh, PA 15260\vspace{0.2cm}}
\affiliation{Theory Group, Fermilab, 
	P.O. Box 500, 
	Batavia, IL 
	60510\vspace{0.2cm}}

\author{Thomas Mehen\footnote{Electronic address: mehen@phy.duke.edu}}
\affiliation{Department of Physics, Duke University, Durham,  NC 27708\vspace{0.2cm}}
\affiliation{Jefferson Laboratory, 12000 Jefferson Ave., Newport News, VA 23606\vspace{0.2cm}}

\date{\today\\ \vspace{1cm} }



\begin{abstract}

Recent observations of the spectrum of $J/\psi$ produced in $e^+e^-$ collisions at the 
$\Upsilon(4S)$ resonance are in conflict  with 
fixed-order calculations using the Non-Relativistic QCD (NRQCD) effective field theory. One
problem is that leading order color-octet
mechanisms predict an enhancement of  the cross section for $J/\psi$ with maximal energy 
that is not observed in the data. However, in this
region of phase space large perturbative corrections (Sudakov logarithms) 
as well as enhanced nonperturbative effects are important. In
this paper we use the newly developed Soft-Collinear Effective 
Theory (SCET) to systematically include these effects. We find that these
corrections significantly broaden the color-octet contribution to the $J/\psi$ 
spectrum. Our calculation employs a one-stage
renormalization group evolution rather than the two-stage evolution used 
in previous SCET calculations. We give a simple argument for why
the two methods yield identical results to lowest order in the SCET power counting.

\end{abstract}

\maketitle

\newpage

\section{Introduction}

Bound states of heavy quarks and antiquarks have been of great interest 
since the discovery of the $\jpsi$~\cite{psi}. In particular
the production of quarkonium is an interesting probe of both perturbative 
and nonperturbative aspects of QCD dynamics.
Production  requires the creation of a heavy  $Q\bar{Q}$ pair with energy 
greater than $2 m_Q$, a scale at which  the
strong coupling constant is small enough that perturbation theory can be used. 
However, hadronization
probes much smaller mass scales of order $m_Q v^2$, where $v$ is the typical velocity
of the quarks in the quarkonium. For $J/\psi$, $m_Q v^2$ is numerically of order $\Lambda_{\rm QCD}$ so the 
production  process is sensitive to nonperturbative physics as well.

A systematic theoretical framework for handling the different scales characterizing both the decay and production of quarkonium is
Non-Relativistic Quantum Chromodynamics (NRQCD)~\cite{Bodwin:1995jh, Luke:2000kz}. NRQCD solves important conceptual as well as
phenomenological problems in quarkonium theory. For instance,  perturbative calculations of the inclusive decay rates for $\chi_c$ mesons
in the color-singlet model suffer from nonfactorizable infrared  divergences~\cite{Barbieri:1976fp}.  NRQCD provides a generalized
factorization theorem  that includes nonperturbative corrections to the color-singlet model, including color-octet decay mechanisms. All
infrared divergences can be factored into nonperturbative matrix elements, so that  infrared safe calculations of inclusive decay rates are
possible~\cite{Bodwin:1992ye}. In addition, color-octet production mechanisms are critical for understanding the production of $J/\psi$ at large transverse
momentum, $p_\perp$, at the Fermilab Tevatron \cite{tevatron}.  NRQCD has been applied to the production and decay of quarkonium in various
experimental settings. However, there are still many challenging problems in quarkonium physics that remain to be
solved~\cite{Bodwin:2002mr}.  One important problem  is the polarization of $J/\psi$ at the Tevatron. NRQCD predicts the $J/\psi$ should
become transversely polarized as the $p_\perp$ of the $J/\psi$ becomes much larger than $2 m_c$~\cite{psipol}. The theoretical prediction
is consistent with the experimental data at intermediate  $p_\perp$, but at the largest measured values of $p_\perp$ the $J/\psi$ is
observed to be slightly longitudinally polarized. At these $p_\perp$, discrepancies at the 3$\sigma$ level are seen in both prompt $J/\psi$
and $\psi^\prime$ polarization measurements~\cite{poldata}.   

New problems have arisen as a result of recent measurements of the spectra of $J/\psi$ produced at the $\Upsilon(4S)$ resonance in $e^+e^-$
collisions by the BaBar and Belle experiments~\cite{Abe:2001za, Aubert:2001pd}.   Leading order NRQCD calculations predict that for most of
the range of allowed energies prompt $J/\psi$ production should be dominated by color-singlet production mechanisms, while color-octet
contributions dominate when the $J/\psi$ energy is nearly maximal. Furthermore, as pointed out in  Ref.~\cite{Braaten:1995ez}, color-octet
processes predict a dramatically different  angular distribution for the $J/\psi$. Writing the differential cross section as 
\begin{eqnarray}
\frac{d \sigma}{d p_\psi \, d \cos\theta} = S(p_\psi)(1+ A(p_\psi) \cos^2 \theta) \, ,
\end{eqnarray}
where $p_\psi$ is the $J/\psi$ momentum and $\cos\theta$ is the angle of the $J/\psi$ with respect to the axis defined by the $e^+e^-$ beams, one finds
the color-singlet mechanism gives $A(p_\psi)\approx 0$ except for large $p_\psi$, where $A(p_\psi)$ becomes large and negative. On the other hand,
color-octet production predicts $A(p_\psi) \approx 1$. The significant enhancement of the  cross section accompanied by the change in angular
distribution were proposed as a distinctive signal of color-octet mechanisms in  Ref.~\cite{Braaten:1995ez}. It was expected that these effects would be
confined to $J/\psi$ whose momentum is within a few hundred  MeV of the maximum allowed.

Experimental results do not agree with these expectations. The cross section data as a function of momentum is binned in intervals of 300 or 500
MeV, depending on the experiment, and the data does not exhibit any enhancement in the bins closest to the endpoint. However, the total cross
section measured by the two experiments exceeds predictions based on the color-singlet model alone. The total prompt $J/\psi$ cross section, which
includes feeddown from $\psi^\prime$ and $\chi_c$ states but not from $B$ decays, is measured to be $\sigma_{tot} = 2.52 \pm 0.21 \pm 0.21$ pb by
BaBar, while Belle measures $\sigma_{tot} = 1.47 \pm 0.10 \pm 0.13$ pb. Estimates of the color-singlet contribution range from $0.4 -0.9$ pb
\cite{Cho:1996cg,Yuan:1996ep,Baek:1998yf,Schuler:1998az}. Furthermore,  $A(p_\psi)$ is measured to be consistent with $1$ (with large errors) for
$p_\psi > 2.6 \,{\rm GeV}$ (Belle)  and $p_\psi > 3.5 \,{\rm GeV}$ (BaBar). These aspects of the data suggest that there is a  substantial
color-octet contribution which is not confined to the very endpoint of the momentum  spectrum but spread over a substantially broader range of
momentum. 

In this paper, we will argue that higher order perturbative and nonperturbative corrections that are enhanced near the endpoint give rise to a broad color-octet
contribution.  The calculation depends on a nonperturbative function, called a shape function,  which parametrizes the distribution of lightcone momentum of the
$\jpsi$ carried by the color-octet $c\bar{c}$ pair produced in the short-distance process. Since the shape function is nonperturbative, our calculation is not
predictive. However, moments of the shape function are NRQCD operators whose size is constrained by the velocity scaling rules of NRQCD. Choosing a simple ansatz
for the shape function whose moments are consistent with velocity scaling rules, we find that the combined perturbative and nonperturbative effects lead to
substantial broadening of the color-octet spectrum in a manner that is consistent with data. Since the shape function that appears in this  calculation also
appears in other processes, it could be extracted from $e^+e^-$ data and used to make predictions for photoproduction and electroproduction once resummed
calculations for these processes become available. Ref.~\cite{Beneke:1999gq} presents a calculation of $J/\psi$ photoproduction which includes a nonperturbative
shape function but not the resummation of large perturbative corrections. An important point of this paper is that nonperturbative and perturbative corrections
combine to produce a spectrum that is broader than that  obtained when only one of the corrections is included. Therefore it would be interesting to revisit
the calculation of photoproduction using the methods of this paper. 

While the calculations of this paper show that the leading color-octet contribution is broad enough to 
be compatible with the  observed $p_\psi$ distributions, other features of the $e^+e^-$ data remain puzzling. In particular, Belle reports a
large cross section for $J/\psi$ produced along with open charm~\cite{Abe:2002rb}:
\begin{eqnarray}
\frac{\sigma (e^+e^- \rightarrow J/\psi c \bar{c})}{\sigma (e^+e^- \rightarrow J/\psi X)} =0.59^{+0.15}_{-0.13}\pm 0.12 \nonumber \, .
\end{eqnarray}
The predicted ratio from leading order color-singlet production mechanisms alone is about $0.2$ \cite{Cho:1996cg,Baek:1998yf} 
 and a large color-octet contribution makes this ratio even smaller. 
In addition to the inclusive measurements, Belle reports a cross section for exclusive double charmonium 
production which exceeds previous theoretical estimates. Recent attempts to address the latter problem can be found in Ref.~\cite{double}.

The experiments also measure the polarization of the $J/\psi$, which can provide important information about the production mechanism. Unfortunately the
current experimental situation is unclear. Polarization is studied by measuring 
\bea
\alpha = \frac{1 - 3 \eta_L}{1 + \eta_L} \, , \nn
\eea
where $\eta_L$ is the fraction of $J/\psi$ which are longitudinally polarized. Both BaBar and Belle measure $\alpha \approx -0.5$ for $p_\psi < 3.5$ GeV.
However, for $p_\psi > 3.5$ GeV  BaBar measures $\alpha= -0.8\pm 0.09$, corresponding to almost completely longitudinally polarized $J/\psi$, while Belle
measures  $\alpha = -0.2\pm 0.2$, which is consistent with  no polarization. Neither measurement attempts to correct for  feeddown effects on the
polarization of the $J/\psi$. Measurements of the polarization  of directly produced $J/\psi$ can discriminate between various production
mechanisms~\cite{Baek:1998yf}. The color-singlet process $e^+e^- \rightarrow J/\psi \overline c c$ produces $J/\psi$  with $\alpha$ rising from zero at
$p_\psi=0$ to 1 at the kinematic endpoint for this process. The  color-singlet process $e^+e^- \rightarrow J/\psi  g g$ prefers longitudinally polarized
$J/\psi$  with $\alpha$ decreasing to almost $-1$ at the kinematic endpoint. Finally,  the leading order color-octet diagrams produce $J/\psi$ with 
$\alpha$ between 0 and -0.07, depending on the relative importance of $S$-wave and $P$-wave production mechanisms.

The NRQCD factorization formalism shows that the differential $\jpsi$ cross section can be written as
\begin{equation} \label{NRQCDprod} 
d \sigma (e^+ e^- \rightarrow \jpsi + X) = \sum_n d
\hat{\sigma} (e^+ e^- \rightarrow c \bar{c}[n]+ X) \langle {\cal O}^{\jpsi}_n \rangle \,,
\end{equation}
where $d \hat{\sigma}$ is the inclusive cross section for producing a
$c\bar{c}$ pair in a color and angular momentum state labeled by $[n] = {}^{2S+1}L_J^{(i)}$. In this notation,
the spectroscopic notation for angular momentum quantum numbers is standard and
$i = 1 (8)$ for color-singlet(-octet) production matrix elements.
The short-distance coefficients are
calculable in a perturbation series in $\alpha_s$. The
long-distance matrix elements $\langle {\cal O}^{\jpsi}_n \rangle$ are
vacuum matrix elements of four-fermion operators in
NRQCD~\cite{Bodwin:1995jh}. These matrix elements scale as some power
of the relative velocity $v \ll 1$ of the $c$ and $\bar{c}$ quarks as
given by the NRQCD power-counting rules. 

\begin{figure}
\begin{center}
\picspace
\includegraphics[width=5in]{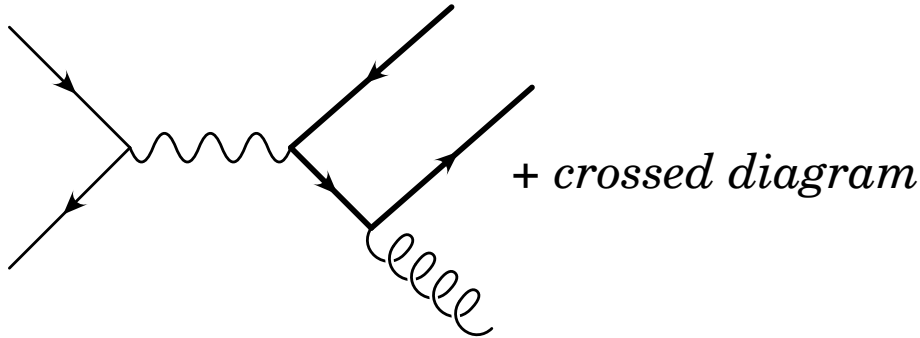}
\picspace
\caption{\it  Leading order short distance amplitudes  for $e^+ e^-\to c\bar{c}  + g$ in QCD.  
\label{NRQCDfig}}
\end{center}
\end{figure}

At lowest order in $v$ the only term in Eq.~(\ref{NRQCDprod}) is the color-singlet
contribution, $[n]={}^3S_1^{(1)}$, which scales as $v^3$.  The  coefficient for this
contribution starts at $O(\alpha^2_s)$ \cite{csrate}. 
Away from the kinematic endpoint $\emax = (s+\mpsi^2)/ (2 \sqrt{s})$,
where $s$ is the center-of-mass energy squared, color-octet contributions also start
at $O(\alpha_s^2)$. Since the color-octet contributions are suppressed by $v^4 \approx 0.1$ 
relative to the leading color-singlet contribution, they are negligible 
throughout most of the allowed phase-space at leading order in perturbation theory.
However, as pointed out in Ref.~\cite{Braaten:1995ez}, there is an $O(\alpha_s)$ contribution 
to color-octet production from the diagrams shown in Fig.~\ref{NRQCDfig}.
In this process the $c\bar{c}$ are in either a ${}^1S_0^{(8)}$ or ${}^3P_J^{(8)}$ configuration. 
The resulting cross sections are proportional to a $\delta-$function which forces the 
energy of the $c\bar{c}$ pair to be maximal \cite{Braaten:1995ez}:
\begin{eqnarray}\label{octet}
\frac{d \hat{\sigma}[^1S_0^{(8)}]}{d z \, d \cos{\theta}} & = & \delta(1-z) 
\frac{4 \pi^2 \alpha^2 \alpha_s e^2_c}{s^2 m_c}
(1-r) (1+\cos^2{\theta}),
\\
\frac{d \hat{\sigma}[{}^3P_0^{(8)}]}{d z \, d \cos{\theta}} & = & \delta(1-z) 
\frac{8 \pi^2 \alpha^2 \alpha_s e^2_c}{s^2 m_c^3} 
\frac{(3+6 r+7r^2)+(3-10 r+7r^2)\cos^2{\theta}}{2 (1-r)} \,, \nn
\end{eqnarray}
where $r = 4 m^2_c /s$, and $z=E_{c\bar{c}}/E^{max}_{c\bar{c}}$ with $E^{max}_{c\bar{c}} = \sqrt{s}(1+r)/2$. 
Here we have used $ \langle {\cal O}^\psi_8(^3P_J)\rangle = (2J+1) \langle {\cal O}^\psi_8(^3P_0)\rangle$
and combined  the short-distance cross sections for the three $P$-wave contributions.
For comparison, in this region of phase space the
color-singlet contribution approaches a constant
\begin{equation}
\lim_{z\rightarrow 1}\frac{d \hat{\sigma}[{}^3S_1^{(1)}]}{d z \, d \cos{\theta}} =
\frac{64 \pi \alpha^2 \alpha^2_s e^2_c}{27 s^2 m_c}(1+r)
\Bigg( \frac{1+r}{1-r} - \cos^2{\theta} \Bigg) \,.
\end{equation}
We see that the $v^4$ suppression of the color-octet contribution is compensated by one power of $\alpha_s$. In addition, the numerical coefficient of the
color-octet contributions is bigger by a factor of $\approx 4 \pi$ which is due to the difference between two- and three-body phase space. Since the color-octet
contribution is proportional  to a singular distribution, sensible comparison with data requires that we integrate the cross section over a region in $z$ of
finite size. It is important to know what range of $z$ must be  integrated over  in order for the leading order calculation to be reliable.
Ref.~\cite{Beneke:1997qw} has identified a class of nonperturbative corrections which are suppressed by powers of $v^2$ but are enhanced by factors of $1/(1-z)$,
so we must integrate from a lower limit no larger than  $z_{min} \sim 1-v^2$. For $z_{min} \sim 1- v^2$ , the integrated color-singlet cross section gets an
additional $v^2$ suppression relative to the color-octet contribution so the ratio of the integrated color-octet to integrated color-singlet is $\sim 4 \pi
v^2/\alpha_s$ which is roughly a factor of 10.  

Note that because $v^2 \approx 0.3$ for $J/\psi$, the cross section must be integrated over a substantial region. For production at the $\Upsilon(4S)$,
$E_{max}= 5.74$ GeV, corresponding to $p_{max}= 4.84$ GeV, so we must integrate above $E \approx  4.0$ GeV, which corresponds to  $p_\psi  \approx 2.5$ GeV.
These numbers show that the spectrum reported in Refs.~\cite{Abe:2001za, Aubert:2001pd}, which plot the data as a function of $p_\psi$, cannot be directly
compared to the leading order NRQCD calculation. Nonperturbative effects smear the color-octet contribution over nearly half the range of allowed
momenta, so the color-octet contribution is not confined to the last momentum bin.  Higher order perturbative corrections are also important. The
next-to-leading order radiative correction to color-octet production has a contribution proportional  to $\alpha_s \log(1-z)/(1-z)$, so perturbative
corrections require a resummation when $1-z \sim \alpha_s$. Since $\alpha_s \sim v^2$,  similar conclusions regarding the reliability of the leading
order calculation arise from  purely perturbative considerations. 

Large perturbative and nonperturbative corrections signal that the NRQCD factorization theorem of
Eq.~(\ref{NRQCDprod}) is no longer valid. The problem is that near the endpoint the $\jpsi$ is recoiling against a gluon jet with energy of
order $M_\Upsilon$ but mass of order  $M_\Upsilon \sqrt{ \Lambda_{\rm QCD}/M_\psi}$. These degrees of freedom are  integrated out of NRQCD, but when we probe the
endpoint region they must be kept as explicit degrees of freedom. The effective theory which correctly describes this kinematic regime is a
combination of NRQCD for the heavy degrees of freedom, and the Soft-Collinear Effective
Theory (SCET)~\cite{Bauer:2001ew,Bauer:2001yr,Bauer:2001ct,Bauer:2001yt} for the light energetic degrees of freedom. The factorization theorem
derived using NRQCD and SCET  will include a nonperturbative  distribution that incorporates the $v^{2n}/(1-z)^n$  nonperturbative corrections
to all orders. In addition, renormalization group equations of SCET can be used to resum large perturbative corrections. A similar treatment of
nonperturbative and perturbative endpoint corrections to the color-octet contributions in the inclusive decay $\Upsilon \rightarrow X +\gamma$
can be found in Ref.~\cite{Bauer:2001rh}.

\section{Factorization}\label{fact}

In this section, we present the factorization theorem for $J/\psi$ production near the kinematic endpoint.
We begin by studying the kinematics of the process to determine when NRQCD breaks down and SCET must be used instead. 
In the $e^+e^-$ center-of-mass (COM) frame, the $c \bar{c}$ pair has momentum
$p_{c\bar c}^\mu = Mv^\mu+\ell^\mu$, where $M=2 m_c$, $\ell^\mu$ is the
residual momentum of the $c\bar{c}$ pair and the four-velocity of the $J/\psi$ is
\begin{equation}
v^\mu = \frac{1}{2} \Bigg( \frac{M_{\psi}}{x \sqrt{s}} n^\mu +  \frac{x \sqrt{s}}{M_{\psi}} \bn^\mu  \Bigg) \,.
\end{equation}
Here $M_{\psi}$ is the $\jpsi$ mass and  $x  = (E_{\psi}+p_{\psi})/\sqrt{s}$. The residual momentum arises because  the $c \bar c$ pair is produced
along with ultrasoft (usoft) quanta which carry $O(\Lambda_{\rm QCD})$ momentum in the rest frame of the $J/\psi$. The lightlike vectors are
$\bn^\mu = (1,0,0,1)$ and $n^\mu=(1,0,0,-1)$, where the $J/\psi$ is moving in the $z$ direction  in the COM frame.  It is
sometimes helpful to write $\ell^\mu$   in terms of the boost that takes one from the $J/\psi$ rest frame to the COM frame:
$\ell^\mu =\Lambda^\mu{}_\nu \hat{\ell}^\nu$. The components of $\hat{\ell}^\mu$  are ${O}(\Lambda_{\rm QCD})$ while the
components of $\ell^\mu$ scale  as: $ \nb \cdot \ell \sim M_\psi\Lambda_{\rm QCD}/(x\sqrt{s}), 
n \cdot \ell \sim x \sqrt{s}\Lambda_{\rm QCD}/M_\psi$ and $  \ell_\perp   \sim  \Lambda_{\rm QCD}$.
The momentum of the virtual photon is $q^\mu = \sqrt{s}/2(n^\mu+\bn^\mu)$ and  the gluon jet has momentum

\begin{equation}\label{jetmom}
p^\mu_X = \frac{\sqrt{s}}{2} \Bigg[ \Bigg(1 - \frac{r}{\hat{x} } \Bigg) n^\mu  +(1-\hat{x}) \bar{n}^\mu \Bigg]  - \ell^\mu \, ,
\end{equation}
where  $\hat{x} = x M /M_\psi$. Away from the endpoint region  ($x \ll 1$) the recoiling gluon jet momentum and invariant mass are both of order
$\sqrt{s}$. Therefore, the jet can be integrated out of the theory and one  obtains the NRQCD factorization formula in Eq.~(\ref{NRQCDprod}).  In this
region $\ell^\mu$ is negligible compared to the large components of $p_X^\mu$ and $p_{c\bar c}^\mu$ and can be set to zero. Therefore the cross section
is not sensitive to motion of the $c\bar c$ within the $J/\psi$. This is evident from the NRQCD factorization formula  which  depends on a single
nonperturbative parameter, $\langle {\cal O}(^1S_0^{(8)}) \rangle$ or $\langle {\cal O}(^3P_0^{(8)}) \rangle$. When  $1-x \sim  \Lambda_{\rm QCD}/M \sim
v^2$, $\sqrt{s}(1-x)$ is the same size as $n \cdot \ell$, so the  residual momentum cannot be neglected. In the endpoint region, the NRQCD factorization 
formula breaks down. A new factorization theorem  is needed which includes a distribution function that parametrizes the nonperturbative
motion of the $c\bar c$ pair within the jet containing the  $J/\psi$. For $1-x = 1/3$, $p_{\psi}= 2.8$ GeV, so the new factorization theorem is relevant
for a significant part of the measured $p_{\psi}$ spectrum.  The failure of NRQCD factorization can also be understood by considering the gluon jet which
was integrated out. When $1-x \sim  \Lambda_{\rm QCD}/M$ the jet is no longer highly virtual:
\begin{eqnarray}
m_X^2 \sim  \Lambda_{\rm QCD} s/M\,.
\end{eqnarray}
Since $m_X^2/E_X^2 \sim \Lambda_{\rm QCD}/M \ll 1$, the gluon jet is composed of energetic particles with small invariant mass
that must be included explicitly in the 
effective theory. 

SCET has collinear degrees of freedom whose momentum scales as  $\nb \cdot p \sim Q$,  $n \cdot p \sim \lambda^2 Q$, and $p^\perp
\sim \lambda Q$. For the process under consideration,  $Q$ is of order $\sqrt{s}$, while $\lambda \sim
\sqrt{1-x}\sim (\Lambda_{\rm QCD}/M)^{1/2}$. SCET also has soft degrees of freedom whose momentum scales as $\lambda$ and  usoft
degrees of freedom whose momentum scales as $\lambda^2$. Heavy quark fields in SCET are the same as in NRQCD when considering quarkonium.  Since a field
redefinition removes the large $O(m_c, m_c v)$ part of their  momentum, derivatives on these fields scale as $\Lambda_{\rm QCD}$ and
therefore they are usoft fields in SCET. Thus the endpoint region of $J/\psi$ production is mediated by SCET  operators involving
usoft (heavy) quark fields and collinear gluon fields.  Soft fields do not enter to the order we are working, so are neglected.


To match onto SCET, matrix elements in QCD are evaluated at the scale $Q$ and expanded in powers of  $\lambda$.  Each order in the
$\lambda$ expansion is reproduced in the effective theory by  a product of Wilson coefficients (which depend only on the large scale
$Q$) and  SCET operators. At each order, one must include all SCET operators which can contribute to the process under consideration, subject to the
restriction that the operators respect the symmetries of the effective  theory. In SCET, gauge transformations can be classified by
their scaling with $\lambda$ just like fields \cite{Bauer:2001yt}.   For $e^+e^- \rightarrow J/\psi + X$,  the relevant operators
must be invariant under collinear and usoft gauge transformations \cite{Bauer:2001yt}. SCET  is not Lorentz invariant, but the
Lorentz invariance of QCD is realized in the form of constraints on the possible operators which appear in the Lagrangian and in
matching calculations. These constraints are called reparametrization invariance (RPI) \cite{Manohar:2002fd}. Usoft and collinear
gauge invariance and RPI  uniquely fix the form of the lowest order operators contributing to $e^+ e^- \rightarrow J/\psi + X$.

In the collinear sector of SCET there is a collinear fermion field $\xi_{n,p}$, a collinear gluon field $A_{n,q}^\mu$, and a
collinear Wilson line
\begin{equation}
W_n(x)=
 \bigg[ \sum_{\rm perms} {\rm exp} 
  \left( -g_s \frac{1}{\bnP} \bn \cdot A_{n,q}(x) \right) \bigg] \,.
\end{equation}
The subscripts on the collinear fields are the lightcone direction
$n^\mu$, and the large components of the lightcone momentum ($\bn\cdot
q, q_\perp$). The operator ${\cal P}^\mu$ projects out the momentum
label~\cite{Bauer:2001ct}.  For example $\bn\cdot {\cal P} \xi_{n,p}
\equiv \bnP \xi_{n,p} = \bn \cdot p\, \xi_{n,p}$.  Likewise in the usoft
sector there is a usoft fermion field $q_{us}$, a usoft gluon field
$A^\mu_{us}$, and a usoft Wilson line $Y$. Using the transformations for each of these
fields under collinear and usoft gauge transformations  given 
in Ref.~\cite{Bauer:2001yt}, we can build invariant operators.
%
%
%
The collinear-gauge invariant field strength is
\begin{equation}\label{colfieldstrength}
G^{\mu\nu}_n \equiv -\frac{i}{g_s} W^\dagger [i{\cal D}_n^\mu 
  + g_sA_{n,q}^\mu, i{\cal D}_n^\nu+g_sA_{n,q'}^\nu ] W ,
\end{equation}
where 
\begin{equation}\label{cov}
i{\cal D}_n^\mu = \frac{n^\mu}2 \bnP + \ppP^\mu + 
\frac{\bn^\mu}2 i n\cdot D,
\end{equation}
and $iD^\mu = i \partial^\mu+g_sA^\mu_{us}$ is the usoft covariant
derivative.  RPI invariance requires the label operators and the usoft
covariant derivatives, which scale  differently with $\lambda$,
to appear in the linear combination appearing in $i{\cal D}_n^\mu$.
The leading piece of $G^{\nu \mu}_n$ is order $\lambda$ and
can be written as $\bn_\nu G^{\nu\mu}_n = i[\bnP ,  B^\mu_\perp]$, where 
\begin{equation}\label{bfield}
B^\mu_\perp =  \frac{1}{g_s} W^\dagger (\ppP^\mu + g_s (A^\mu_{n,q})_\perp)W. 
\end{equation}
The subscript $\perp$ on $B_\perp^\mu$ indicates that $\mu$  must be a 
perpendicular direction. The leading operator for $e^+e^- \rightarrow c\bar c g$
must have a $(A^\mu_{n,q})_\perp$ field operator to create
the collinear gluon in the final state. $B^\mu_\perp$ is the collinear gauge invariant 
generalization of this field. 

The $O(\lambda)$ operator which creates a heavy quark and antiquark 
in a ${}^1S_0$ state with a collinear gluon is 
\begin{equation} \label{1s0op1}
   \psi^\dagger_{{\bf p}} \, \Gamma^{(8,{}^1S_0)}_{\alpha \mu}(- n\cdot v\,\bnP;\mu)   B^\alpha_\perp \chi_{-\bf p} \,.
\end{equation}
This is collinear gauge invariant because $B_\perp^\mu$ is invariant and the heavy
quark fields do not transform under collinear gauge transformation. It is also  usoft gauge invariant
since
\begin{equation}  
   \chi_{-{\bf p}}\rightarrow V_{\rm us} \chi_{- {\bf p}},
   \qquad  \psi_{\bf p} \rightarrow V_{\rm us} \psi_{\bf p},
   \qquad  B^\alpha_\perp \rightarrow V_{\rm us}B^\alpha_\perp V^\dagger_{\rm us} \, ,
\end{equation}
under usoft gauge transformations. Type-III reparametrization invariance \cite{Manohar:2002fd} requires $\bnP$  to appear multiplied  by
$n\cdot v$ \cite{Pirjol:2002km}. The operator is not invariant under type-I and -II  transformations but can be made invariant by adding $\lambda$
suppressed operators which  are not needed at the order we are working. The factor of $-\bnP$ in Eq.~(\ref{1s0op1})  gives the large component
of the momentum of the jet at the endpoint. This is the part of $\bn \cdot p_X$ that survives as $\Lambda_{\rm QCD}/M$ and 
$1-x \rightarrow 0$, so that $-n\cdot v\,\bnP = s(1-r)/M$.  At leading order, the function $\Gamma^{(8,{}^1S_0)}_{\alpha \mu}$ is determined by requiring the SCET
matrix element of Eq.~(\ref{1s0op1}) to agree with  the lowest order QCD diagrams for $e^+ e^- \rightarrow c \bar c g$, shown in Fig.~\ref{matchingfig}:
\begin{figure}
\begin{center}
\picspace
\includegraphics[width =5in]{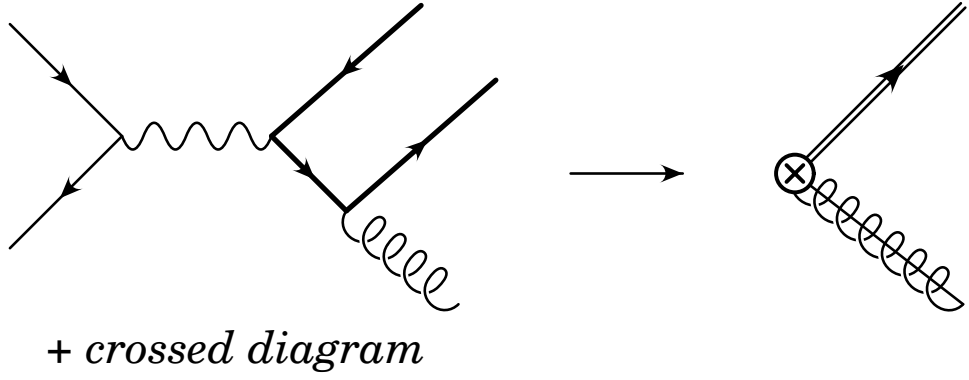}
\picspacehalf
\caption{\it  Matching the production amplitude for $e^+ e^-\to c\bar{c}  + g$ in QCD and SCET.  
Collinear gluons are represented by a spring with a line through it.  
\label{matchingfig}}
\end{center}
\end{figure}
\begin{equation}\label{1s0matching}
 \Gamma^{(8,{}^1S_0)}_{\alpha \mu} 
   = \frac{2 g_s e e_c}{M} \epsilon_{\alpha \mu}^\perp  \,,
\end{equation}
where $\epsilon_{\alpha \mu}^\perp = \epsilon^\perp_{\alpha\mu\rho\beta} \bn^\rho n^\beta/2$.  The leading SCET color-octet ${}^3P_J$ operator is
\begin{equation} \label{3pjop1}
   \psi^\dagger_{{\bf p}} \Gamma^{(8,{}^3P_J)}_{\alpha \mu \sigma \delta}(-n\cdot v \bnP; \mu) 
   B^\alpha_\perp  \Lambda \cdot \widehat{{\bf p}}^\sigma 
   \Lambda \cdot \bsigma^\delta \chi_{-\bf p} \,,
\end{equation}
where $\widehat{{\bf p}} = {\bf p}/M$ and the leading order matching coefficient is
\begin{eqnarray}\label{3p0matching}
\Gamma^{(8,{}^3P_J)}_{\alpha \mu \sigma \delta} &=&
\frac{2 i g_s e e_c}{M}
\bigg\{ 
g^{\perp}_{\alpha \sigma} \bigg[ 2\bigg( \frac{1+r}{1-r}\bigg) g_{\mu\delta}-
        \frac{1}{r} (1-r)  \bn_\delta n_\mu \bigg]
\nn \\
&& \phantom{  4 g_s e e_b  \bigg\{ }
-g^{\perp}_{\alpha \delta} \bigg[ 2 g_{\mu \sigma}-
        \frac{1}{r}( 1+ r) n_\mu \bn_\sigma \bigg] 
 - g^{\perp}_{\alpha \mu} (n_\delta \bn_\sigma+\bn_\delta n_\sigma) \bigg\},
\end{eqnarray}
where $g^\perp_{\mu\nu} = g_{\mu\nu} - (n_\mu \bn_\nu + n_\nu \bn_\mu)/2$.

It is possible to show that the rate $e^+e^-\to \jpsi + X$ in the endpoint region can be factored into a hard coefficient $\sigma_0$, a
collinear jet function $J$, and a usoft function $S$, as
\begin{equation}\label{factformula}
\frac{d\sigma^{[n]}}{dz} = P[r,z] \sigma_0^{[n]}
\int_z^{1} d\xi \, S^{[n]}(\xi) \, J(s(1+r)(\xi-z)),
\end{equation}
where $z = E_\psi/E^{max}_{\psi}$, and 
\begin{equation}
P[r,z] = \frac{\sqrt{(1+r)^2 z^2 - 4 r}}{1-r}
\end{equation}
is a phase space factor. Note that $P[r,1] = 1$.  The derivation is given in Appendix~\ref{factproof}. For convenience we have integrated
over $\theta$. The $\theta$ dependence of the resummed cross sections are the same as in Eq.~(\ref{octet}). 


The jet function in Eq.~(\ref{factformula}) is independent of the state of the $c\bar c$ pair in the $\jpsi$, and is defined as
\begin{eqnarray}
J(\nb \cdot p \, n\cdot k +p_\perp^2) 
&=&  \\
& & \hspace{-3ex}  - \frac{s(1+r)}{4 \pi}\,{\rm Im} \left[i \int d^4y \, e^{i k \cdot y} \,
\langle 0 | T \left( {\rm Tr} \left[ T^B B^{(0) \beta}_{\perp}(y) \right] 
{\rm Tr}\left[ T^B B^{(0)}_{\perp\,\beta}(0) \right]  \right) | 0 \rangle  \right] \nonumber ,
\end{eqnarray}
where the factor $s(1+r)$ is chosen to provide a convenient normalization for the  process we are interested in. $J(\bn \cdot p \, n\cdot
k +p_\perp^2)$ is a function of the label momenta  $\bn \cdot p$ and $p_\perp^2$, and the usoft momentum $n\cdot k$, which follows from the
collinear Lagrangian containing only the $n\cdot \partial$ derivative~\cite{Bauer:2001yt}. Furthermore RPI requires that the jet function depends on the
combination $\nb \cdot p \, n\cdot k +p_\perp^2$. Here we parameterize the jet momentum so that $\nb \cdot p = -\nb\cdot p_X$
and $p_\perp = 0$. The jet function is perturbatively calculable since it is determined by physics at the scale $\sqrt{s \Lambda_{\rm
QCD}/m_c}\gg \Lambda_{\rm QCD}$. The $O(\alpha_s)$ calculation of the jet function, as well as the derivation of the leading order renormalization group  equations,
is given in Appendix \ref{jetapp}. 

The shape functions, defined in terms of usoft fields, depend on the state of the $c\bar c$ pair, so there are different
shape functions for the $^1S_0^{(8)}$ and $^3P_J^{(8)}$ contributions.  They are
\begin{eqnarray}\label{formalshapeS}
S^{(8,^1S_0)}(\ell^+) &=& \frac{
\langle 0 | \chi^\dagger_{-{\bf p}} T^B \psi_{\bf p} 
\,a_\psi^\dagger a_\psi\, \delta(\ell^+ - i n\cdot D)
\psi^\dagger_{\bf p} T^B \chi_{-{\bf p}} | 0\rangle
}{4 m_c\langle {\cal O}^\psi_8(^1S_0)\rangle},\\
\label{formalshapeP}
S^{(8,^3P_0)}(\ell^+) &=& \frac{
\langle 0 | \chi^\dagger_{-{\bf p}}\left(-\frac{i}{2} D\cdot \sigma T^B\right) \psi_{\bf p} 
\,a_\psi^\dagger a_\psi\, \delta(\ell^+ - i n\cdot D)
\psi^\dagger_{\bf p} \left(-\frac{i}{2} D\cdot \sigma T^B\right) \chi_{-{\bf p}} | 0\rangle
}{4 m_c\langle {\cal O}^\psi_8(^3P_0)\rangle},
\end{eqnarray}
where $\ell^+ = n\cdot \ell$. Note that the combination $d\ell^+ S^{[n]}(\ell^+)$ is boost invariant, so $d\ell^+ S^{[n]}(\ell^+)  = d\hat\ell^+
S^{[n]}(\hat\ell^+)$, where $\hat\ell^+$ is the residual momentum in the $\jpsi$ rest frame.  It is useful to go to the $\jpsi$ rest frame  when evaluating the shape
functions, since in that frame derivatives acting on the heavy quark fields scale as $m_cv^2$. The normalizations of these functions are defined so that $\int
d\ell^+ S^{(8,^{2S+1}L_J)}(\ell^+) =1$.   The jet function normalization is chosen so that the hard coefficients correspond to the lowest order total cross section
for the color-octet contributions:

\begin{eqnarray}\label{hardS}
\sigma_0^{(8,^1S_0)} &=& \frac{32\pi^2\alpha_s\alpha^2 e_c^2}{3s^2} 
\frac{\langle {\cal O}^\psi_8(^1S_0)\rangle}{m_c} (1 - r), \\
\label{hardP}
\sigma_0^{(8,^3P_0)} &=& \frac{32\pi^2\alpha_s\alpha^2 e_c^2}{3s^2} 
\frac{\langle {\cal O}^\psi_8(^3P_0)\rangle}{m_c^3} \frac{3 + 2 r + 7 r^2}{1 - r} .
\end{eqnarray}

To see how the factorization theorem in Eq.~(\ref{factformula}) reduces to the leading
order cross section in the appropriate limit, we use the result in Appendix \ref{jetapp} for
the lowest order jet function,
\begin{eqnarray}
\label{leadingjet}
J(\bn\cdot p\, n\cdot k + p_\perp^2) = \delta(\xi-z)\,,
\end{eqnarray}
where $\xi$ is defined by
\begin{eqnarray}
p_X^2 &=& s(1+r)\left[1 - z + \frac{1-r}{1+r}\left(\frac{\bar\Lambda}{M} - \frac{\hat\ell^+}{M}
\right)\right]\nonumber\\
&\equiv& s(1+r)(\xi-z),
\end{eqnarray}
and $\bar\Lambda = M_\psi-M$.
Then we combine Eqs.~(\ref{factformula}),  (\ref{formalshapeS}) and (\ref{leadingjet}), where $S^{[n]}$ as a function of $\xi$ is defined as
\begin{equation}
S^{[n]}(\xi)\equiv M\frac{1+r}{1-r} S^{[n]}(\hat\ell^+),
\end{equation}
so that $\int d\xi S^{[n]}(\xi) = 1$, to obtain
\begin{equation}\label{dsdzs}
\frac{d\sigma^{(8,^1S_0)}}{dz} = \frac{P[r,z]\sigma_0^{(8,^1S_0)}}{4 m_c\langle {\cal O}^\psi_8(^1S_0)\rangle}
\langle 0 | \chi^\dagger_{-{\bf p}} T^B \psi_{\bf p} 
\,a_\psi^\dagger a_\psi\, \delta\left[1-z - \left(\frac{1-r}{1+r}\right)\frac{i n\cdot \hat D-\bar\Lambda}{M}\right]
\psi^\dagger_{\bf p} T^B \chi_{-{\bf p}} | 0\rangle
\end{equation}
for the $^1S_0^{(8)}$ contribution, for example. Here, $n\cdot \hat D  = M_\psi/(x\sqrt{s}) \, n\cdot D$ and scales like $m_c v^2$
or $\Lambda_{\rm QCD}$.
 When $1-z \gg (i n\cdot \hat D-\bar\Lambda)/M$, we can drop the derivatives inside the $\delta$-function and
pull the $\delta$-function outside the matrix element. The final result is $d\sigma^{(8,^{2S+1}L_J)}/dz =\sigma_0^{(8,^{2S+1}L_J)} \delta(1-z)$,
which agrees with Eq.~(\ref{octet}).

\section{Resumming Sudakov Logarithms}

The leading order color-octet contribution is proportional to $\delta(1-z)$.  The next-to-leading order radiative corrections have
contributions of the form $\alpha_s \ln(1-z)/(1-z)$.  Clearly  when $z \sim 1-\alpha_s$ these corrections are large and must be resummed. This can be
accomplished in a straightforward manner by using the renormalization group equations of SCET. The resummation is most easily carried out by taking
moments  with respect to $z$, then the large corrections as $z\rightarrow 1$ become large logs of $N$ in the expression for the $N$th moment. 
In this section, the resummation for the color-octet ${}^1S_0$ production mechanism will be described explicitly.  The moments of the cross section in
Eq.~(\ref{factformula}) are 
\begin{eqnarray}\label{moment}
\int_0^1dz z^N \frac{d\sigma^{(8,^1S_0)}}{dz} &=& \sigma_0^{(8,^1S_0)}(\mu) 
\int_0^1dz\,z^N \int_z^1 d\xi \,S^{(8,^1S_0)}(\xi,\mu) \, J(s(1+r)(\xi-z),\mu) \\
&=& \sigma_0^{(8,^1S_0)}(\mu) \int_0^1 d\xi \int_0^\xi dz \, z^N \, S^{(8,^1S_0)}(\xi,\mu) \, J(s(1+r)(\xi-z),\mu) \nonumber \\
&=& \sigma_0^{(8,^1S_0)}(\mu) \int_0^1 d\xi \, \xi^{N+1}\int_0^1 du \,u^N \,S^{(8,^1S_0)}(\xi,\mu) \, J(s(1+r)\xi(1-u),\mu) .
\nonumber 
\end{eqnarray}
In the last line of  Eq.~(\ref{moment}), we have made the substitution $z = u \, \xi$. Since the large logs come from the region $\xi, z \approx 1$, 
$P[r,z]$ can be replaced with $ P[r,1] =1 $ and the factor of $\xi$ in the argument of the jet function can be set equal to 1. Then  the moments factorize:
\begin{eqnarray}\label{prod}
\sigma^{(8,^1S_0)}_N = \sigma_0^{(8^1S_0)}(\mu) \, S_N^{(8,^1S_0)}(\mu) \, J_N(\mu) \, ,
\end{eqnarray}
where  
\begin{eqnarray}\label{moments}
\sigma^{(8,^1S_0)}_N &=& \int_0^1dz \, z^N \, \frac{d\sigma^{(8,^1S_0)}}{dz}   \, , \\
 S_N^{(8,^1S_0)}(\mu) &=& \int_0^1d\xi \, \xi^N \, S^{(8,^1S_0)}(\xi,\mu) \, , \nonumber \\
 J_N(\mu) &=& \int_0^1 du \,u^N \,J(s(1+r)(1-u),\mu) \nonumber \, .
\end{eqnarray}
Note that we are only interested in the large $N$ moments, so have used $S_{N+1} = S_N + O(1/N)$.

To resum logarithms we must find the renormalization group equations for the three terms on the right hand side 
of Eq.~(\ref{prod}). $\sigma_0^{(8,^1S_0)}(\mu)$ is proportional to the square of the matching coefficient 
for the lowest order ${}^1S_0^{(8)}$ operator Eq.~(\ref{1s0op1}), so its anomalous dimension is simply twice the anomalous dimension
of that operator, which is calculated in Ref.~\cite{Bauer:2001rh}:
\begin{eqnarray}\label{hrge}
\mu \frac{d}{d \mu} \sigma_0^{(8^1S_0)}(\mu) = 
\left[  -\frac{2 C_A \alpha_s}{\pi}\log\left( \frac{M \,\bar \mu }
{s (1-r)}\right) - \frac{2\alpha_s}{\pi}\left(\frac{17}{12}C_A - \frac{n_f}{6}\right) 
\right] \sigma_0^{(8^1S_0)}(\mu) \, ,
\end{eqnarray}
where ${\bar \mu}^2 = 4\pi e^{-\gamma}\mu^2$. $\sigma_0^{(8,^3P_0)}(\mu)$ has the same anomalous dimension.  
The scale appearing in the logarithm is $-n\cdot v\bnP = s(1-r)/M$.

The renormalization group equations for the moments of the jet function are calculated in Appendix~\ref{jetapp}. The result is
\begin{eqnarray}\label{jrge}
\mu \frac{d}{d\mu} J_N(\mu) = \left[ \frac{2 C_A \alpha_s}{\pi}\log\left( \frac{\bar \mu^2}
{s(1+r)} \frac{N}{N_0}\right) +\frac{2\alpha_s}{\pi}\left(\frac{11}{12}C_A - \frac{n_f}{6}\right) 
\right] J_N(\mu) \, ,
\end{eqnarray}
where $N_0=e^{-\gamma}$. From Eq.~(\ref{moments}) it is clear that the only dimensional quantity  appearing in the definition 
of $J_N$ is $s(1+r)$, so this is the scale that appears in the logarithm of the anomalous dimension 
in Eq.~(\ref{jrge}).

Since  the moments $\sigma^{(8,^1S_0)}_N$ are  physical and therefore 
$\mu$ independent, the renormalization group equation for 
$S_N$ immediately follows: 
\begin{eqnarray}\label{srge}
\mu \frac{d}{d\mu} S_N(\mu) = \left[ -\frac{2 C_A \alpha_s}{\pi}\log\left( \frac{\bar \mu}{M}
\frac{1-r}{1+r} \frac{N}{N_0}\right) + \frac{\alpha_s C_A}{\pi} \right] S_N(\mu)\, .
\end{eqnarray}
The anomalous dimension of $S_N$ does not depend on the angular momentum quantum number of the $c\bar c$ state, so the resummation is identical for both
${}^1S_0$ and  ${}^3P_0$ production mechanisms. Note that the scale $M(1+r)/(1-r)$ appearing in Eq.~(\ref{srge}) also appears dividing $i n\cdot \hat D$ in
Eq.~(\ref{dsdzs}).

We are able to obtain the anomalous dimension for $S_N$ because we can calculate the renormalization group equations for the hard coefficient and jet function,
and because the moments of the cross section are $\mu$ independent. Direct evaluation of the anomalous dimension for the shape function from the definition in
Eq.~(\ref{formalshapeS}) is complicated by the projection operator $a_\psi^\dagger a_\psi$. In quarkonium decays there is no such projection operator, and the
derivation of the evolution equation for the decay shape function is straightforward~\cite{Bauer:2001rh,Bauer:2001yt} . The result is remarkably similar to
Eq.~(\ref{srge}). The only difference is that the scale appearing in the logarithm is $2m_Q$ in decay as opposed $M(1+r)/(1-r)$ above. Thus the coefficient of
the logarithm, often referred to as the cusp anomalous dimension, and the second term in the square brackets of  Eq.~(\ref{srge}) is the same in production and
decay. This holds to all orders $\alpha_s$ because both processes obey a similar factorization theorem. Specifically the jet functions are identical in the two
processes, and the hard functions obey the same renormalization group equation since the same SCET operator mediates production and decay. The only differences
between the renormalization group equations for the two processes are the scales appearing in logarithms.

Defining $\mu_H = (s/M)(1-r)$ and $y_0 = r (1+r)/(1-r)^2 (N_0/N)$, we see that the logarithms in the hard, jet and usoft
functions are minimized at the scales $\mu_H, \mu_H \sqrt{y_0}$ and $\mu_H y_0$ respectively. Large logarithms of $N$ 
are resummed by evolving the jet and usoft functions to their respective scales. The evolution  can also be done in one
step by defining separate renormalization scales for collinear and usoft loops. Loops whose momenta scale like
$(1,\lambda^2, \lambda)$ come with a factor of $\mu_c^{4-D}$ and loops whose momenta scales like $(\lambda^2,
\lambda^2,\lambda^2)$ come with a factor $\mu_u^{4-D}$. This idea is similar to the velocity renormalization group in
NRQCD \cite{Luke:2000kz}.  The renormalization group equations for $J_N$ and $S_N$ take the form
\begin{eqnarray}\label{twomu}
\mu_c \frac{d} {d \mu_c} J_N &=& \gamma_J^N(\mu_c) J_N \, ,\\
\mu_u \frac{d} {d \mu_u} S_N &=& \gamma_S^N(\mu_u) S_N \, .\nonumber
\end{eqnarray}
Factorization of usoft and collinear degrees of freedom guarantees that $\gamma_J$ is a function  of $\mu_c$ only and
that $\gamma_S$ is a function of $\mu_u$ only. The scales are however correlated,  so that $\mu_c = \mu_H \sqrt{y}$ and
$\mu_u = \mu_H y$. Evolving the variable $y$ from $1$ to $y_0$ simultaneously resums large logs in both $J_N$
and $S_N$. 

Defining $\tilde{\Gamma}_N =J_N S_N$, the evolution equation for $\tilde{\Gamma}_N$ as function of $y$ is
\begin{eqnarray}\label{yrge}
y \frac{d} {d y} \tilde{\Gamma}_N = \left(\frac{1}{2}\gamma_J^N(\mu_H \sqrt{y}) +\gamma_S^N(\mu_H y)\right)
\tilde{\Gamma}_N \, .
\end{eqnarray}
This equation is easily integrated to obtain  the following expression for the resummed moments:
\begin{equation}
\label{fullyresummed}
\sigma_N^{(8,^1S_0)}= 
\sigma_0^{(8,^1S_0)} S_N^{(8,^1S_0)}(\mu_H y_0)\, e^{\log(N) g_1(\chi) + g_2(\chi)},
\end{equation}
where
\begin{eqnarray}
\label{gis}
g_1(\chi) &=&
-\frac{2 C_A}{\beta_0\chi}\left[(1-2\chi)\log(1-2\chi)
-2(1-\chi)\log(1-\chi)\right], \\
g_2(\chi) &=& -\frac{8 \Gamma^{\rm adj}_2}{\beta_0^2}
  \left[-\log(1-2\chi)+2\log(1-\chi)\right] - \log(1-\chi)\nonumber\\
 && - \frac{2C_A\beta_1}{\beta_0^3}
   \left[\log(1-2\chi)-2\log(1-\chi)
  +\frac12\log^2(1-2\chi)-\log^2(1-\chi)\right] \nonumber\\
 && -
 \frac{2C_A}{\beta_0} \log(1-2\chi) -
\frac{4C_A}{\beta_0}\log \left(r\frac{1+r}{(1-r)^2}N_0\right) \left[\log(1-2\chi)-\log(1-\chi)\right]\,,\nonumber
\end{eqnarray}
$\chi=\log (N)\, \alpha_s(\mu_H)\beta_0/4\pi$, $\Gamma^{\rm adj}_2 = C_A[C_A(67/36 - \pi^2/12) - 5n_f/18]$, $\beta_0 = (11C_A-2n_f)/3$, and
$\beta_1 = (34C_A^2-10C_A n_f-6C_F n_f)/3$.  $\Gamma^{\rm adj}_2$ is the $O(\alpha_s^2)$ piece of the cusp anomalous dimension, which was taken from Ref.~\cite{cusprefs}.

The expression in Eq.~(\ref{fullyresummed}) gives the resummed expression for the moments of the differential cross section to next-to-leading logarithmic
order. To obtain the differential cross section, the inverse-Mellin transform  of Eq.~(\ref{fullyresummed}) must be taken.  Using the results of
Ref.~\cite{Leibovich:1999xf}, we find:
\begin{eqnarray}
\label{zspaceresummed}
\frac{d\sigma^{(8,^1S_0)}}{dz}&=& -\int_z^1 \frac{d\xi}{\xi}\, P[r,z]\,\sigma_0^{(8,^1S_0)}\,S^{(8,^1S_0)}\left(\xi\right)\\
&&\phantom{ -\int_z^1 d\xi\, }
z \frac{d}{dz} \left\{
\theta(\xi-z) \; \frac{\exp [ l g_1[\alpha_s \beta_0 l/(4\pi)] +
g_2[\alpha_s \beta_0 l/(4\pi)]]}{\Gamma[1-g_1[\alpha_s \beta_0
l/(4\pi)] - \alpha_s \beta_0 l/(4\pi) g_1^\prime[\alpha_s \beta_0
l/(4\pi)]]}\right\} \,,\nonumber
\end{eqnarray}
where $l \approx -\log(\xi-z)$, $\alpha_s \equiv \alpha_s(\mu_H)$, and the shape function contains no large logarithms.\footnote{The Landau pole in Eq.~(\ref{zspaceresummed}) should be dealt with in the same fashion as in $B$ decays~\cite{Leibovich:2001ra}.}
To obtain the octet ${}^3P_0$ 
contribution let $\sigma^{(8,^1S_0)} \to \sigma^{(8,^3P_0)}$ and $S^{(8,^1S_0)} \to S^{(8,^3P_0)}$.

In this paper,  the jet function and soft functions are first factorized and then run down to their appropriate scales. By linking the scales through the
introduction of the variable $y$, large logarithms have been resummed in a single step using Eq.~(\ref{yrge}). This approach to renormalization group evolution,
which we will refer to as one-stage running, is identical to  the method employed in Ref.~\cite{ks} for the resummation of $b\rightarrow s \gamma$. 
Early applications of SCET \cite{Bauer:2001ew} adopt a two-stage running approach. First, operators in SCET are evolved from the hard scale, $\mu_H$, to the
intermediate scale, which for the process under consideration is $\mu_H \sqrt{y_0}$. Then collinear degrees of freedom are integrated out, leaving only usoft
degrees of freedom. Nonlocal operators in the usoft effective theory are then further evolved to the scale $\mu_H y_0$. This approach to resummation in the
process $b\rightarrow s \gamma$ \cite{Bauer:2001ew} has been shown to yield identical results at the next-to-leading log accuracy as one-stage running \cite{ks}.
Applications of SCET to quarkonium decay in Refs.~\cite{Bauer:2001rh,Fleming:2002rv,Fleming:2002sr} also use a two-stage approach. 

To see why the  two methods yield equivalent results (to leading order in $\lambda$) we first note that
the two-stage running approach of Ref.~\cite{Bauer:2001rh} leads to the following evolution equations
for $\tilde{\Gamma}_N =J_N S_N$:
\begin{eqnarray}\label{twostage}
\mu \frac{d} {d \mu} \tilde{\Gamma}_N &=& (\gamma_J^N + \gamma_S^N) \tilde{\Gamma}_N \qquad  \qquad \qquad
\mu_H \sqrt{y_0} \leq \mu \leq \mu_H \\
&\equiv& -2 \gamma_H \tilde{\Gamma}_N\  \nonumber \\
\mu \frac{d} {d \mu} \tilde{\Gamma}_N &=& \gamma_S^N \tilde{\Gamma}_N 
\qquad  \qquad \qquad \mu_H \,y_0 \leq \mu \leq \mu_H \sqrt{y_0} 
\, . \nonumber
\nonumber 
\end{eqnarray}
The first stage corresponds to running the coefficient of the SCET operator  from the scale $\mu_H$ down to the scale
$\mu_H \sqrt{y_0}$. Note the evolution in this stage does not depend on the moment $N$. The second corresponds  to the
evolution of the usoft function in the purely usoft theory down to the scale $\mu_H y_0$. The difference in the two-stage
running in Eq.~(\ref{twostage}) and the one stage running in Eq.~(\ref{yrge}) can be visualized geometrically as
integrating Eq.~(\ref{twomu}) along two different paths in the $(\mu_u, \mu_c)$ plane \cite{mss}.  The evolution in
Eq.~(\ref{twostage}) corresponds to the path
\begin{eqnarray}
(\mu_u,\mu_c) &=& \mu_H (y,y)   \qquad \qquad \qquad \sqrt{y_0} < y < 1 \\
              &=& \mu_H (y,\sqrt{y_0}) \qquad \qquad y_0 < y < \sqrt{y_0}  \, ,\nonumber
\end{eqnarray}
while integrating Eq.~(\ref{yrge}) corresponds to the path 
\begin{eqnarray}
(\mu_u,\mu_c) &=& \mu_H (y,\sqrt{y})   \qquad \qquad y_0 < y < 1  \, .
\end{eqnarray}
Since the paths begin and end on the same point the difference can be expressed as an integral over a closed
loop in the $(\mu_u,\mu_c)$ plane, and will be zero if the anomalous dimension vector,
$\vec{\gamma} =(\gamma_S^N,\gamma_J^N)$ has a vanishing curl: $\vec{\nabla} \times \vec{\gamma}=0$
where $\vec{\nabla} = (\mu_u d/d \mu_u, \mu_c d/d \mu_c )$, or equivalently,
\begin{eqnarray}
\mu_c \frac{d}{d \mu_c} \gamma_S^N = \mu_u \frac{d}{d \mu_u}\gamma_J^N \, .
\end{eqnarray}
At leading order in $\lambda$, this equation is trivially satisfied due to factorization, so the two-stage and one-stage evolution will give identical results. 
This may not be true for operators appearing at higher orders in $\lambda$ since the factorization only holds to lowest order in $\lambda$.

\section{Phenomenology}

Before we can investigate the phenomenological consequences of our analysis we must determine the shape function. Unfortunately not much is known about the octet
shape functions in $J/\psi$ production. They also arise in photo- and electroproduction~\cite{Beneke:1997qw}, so once the resummed calculations of these processes
are available the universality of the shape functions could be tested. In this paper we use a model of the shape function to fit the available data so our
calculation is not predictive. However, the shape function is not completely arbitrary because the moments of the shape function are NRQCD operators whose sizes are
constrained by NRQCD power counting rules. For example,  taking
moments of Eq.~(\ref{formalshapeS}) gives:
\begin{eqnarray}
S_N^{(8,^1S_0)} & = &  \int d \hat \ell^+ (\hat \ell^+)^N \, S^{(8,^1S_0)}(\hat \ell^+) \nn \\
         & = & \frac{ \langle  0 | \chi^\dagger_{-{\bf p}'} T^A \psi_{{\bf p}'}
         a^\dagger_\psi a_\psi  (in\cdot \hat D)^N \psi^\dagger_{\bf p} T^A \chi_{-{\bf p}} 
         | 0 \rangle}{4 m_c \langle {\cal O}^\psi_8 (^1S_0) \rangle} \,.
\end{eqnarray}
Each derivative is of order $m_c v^2 \sim \Lambda_{\rm QCD}$, so the $N^{\rm th}$ moment is 
${\cal O} (\Lambda^{N}_{\rm QCD})$. 

For our model of the shape function we adopt a modified version of a model used in the decay of $B$ mesons~\cite{Leibovich:2002ys},
\begin{equation}\label{sfmodel}
f(\hat\ell^+) = \frac{1}{\bar\Lambda} \frac{a^{ab}}{\Gamma (ab)} (x -1)^{ab - 1}
e^{-a(x-1)}  \,, \hspace{10ex} x = \frac{\hat\ell^+}{ \Lambda} \,,
\end{equation}
where $a$ and $b$ are adjustable parameters and $\bar\Lambda= M_\psi - M$ . Note $\hat\ell^+$ is the residual momentum of the $c\bar{c}$ pair in the rest frame of the $J/\psi$. The first three moments of Eq.~(\ref{sfmodel}) are
 \begin{eqnarray}
m_0 = \int_{\bar\Lambda}^\infty d\hat\ell^+\, f(\hat\ell^+) =  1, & &  
m_1 = \int_{\bar\Lambda}^\infty d\hat\ell^+\, \hat\ell^+\, f(\hat\ell^+) =   \bar\Lambda (b+1), \nn \\
m_2 = \int_{\bar\Lambda}^\infty d\hat\ell^+\, (\hat\ell^+)^2\, f(\hat\ell^+) &=& 
   \bar\Lambda^2 \bigg( \frac{b}{a} + (b+1)^2 \bigg) \,.
\end{eqnarray}
Since $\bar\Lambda \sim {\cal O}(\Lambda_{\rm QCD} )$, any choice with  $a \sim b \sim {\cal O}(1)$ gives the desired scaling for the moments.  For the sake of simplicity,
we assume that the parameters in the shape function model are the same for both of the color-octet contributions. The data from both BaBar and Belle include feeddown from
$\psi'$ and $\chi_{cJ}$ so these contributions must be taken into account. Because the $\chi_c$ are $P$-wave states, the cross section for their production is $v^2$
suppressed  relative to $\jpsi$ and the feeddown to $\jpsi$ is further suppressed by relatively small branching fractions. For this reason, we neglect feeddown from $\chi_c$
in our analysis. However, feeddown from $\psi^\prime$ states   is not suppressed and must be included. The shape function for $\psi^\prime$ can be different from that of
$\jpsi$ but for simplicity we will assume the same form for both shape functions. Feeddown will affect  the overall normalization of the color-octet cross section which is
given by the linear combination $\langle {\cal O}^\psi_8 (^1S_0)\rangle + 3.8 \langle {\cal O}^\psi_8 (^3P_0)\rangle/m_c^2$, where $ \langle {\cal O}^\psi_8
(^{2S+1}L_J)\rangle = \langle {\cal O}^{\jpsi}_8 (^{2S+1}L_J)\rangle +{\rm Br}(\psi^\prime \to \jpsi) \langle {\cal O}^{\psi^\prime}_8 (^{2S+1}L_J)\rangle$. Likewise, the
normalization of the color-singlet contribution is given by 
\begin{equation}
\langle {\cal O}^\psi_1 (^3S_1)\rangle = \langle {\cal O}^{\jpsi}_1 (^3S_1)\rangle + {\rm Br}(\psi'\to\jpsi) \langle {\cal O}^{\psi'}_1 (^3S_1)\rangle 
= 1.45 {\rm\ GeV}^3 \,.  
\end{equation}

\begin{figure}
\begin{center}
\includegraphics[width=6.25in]{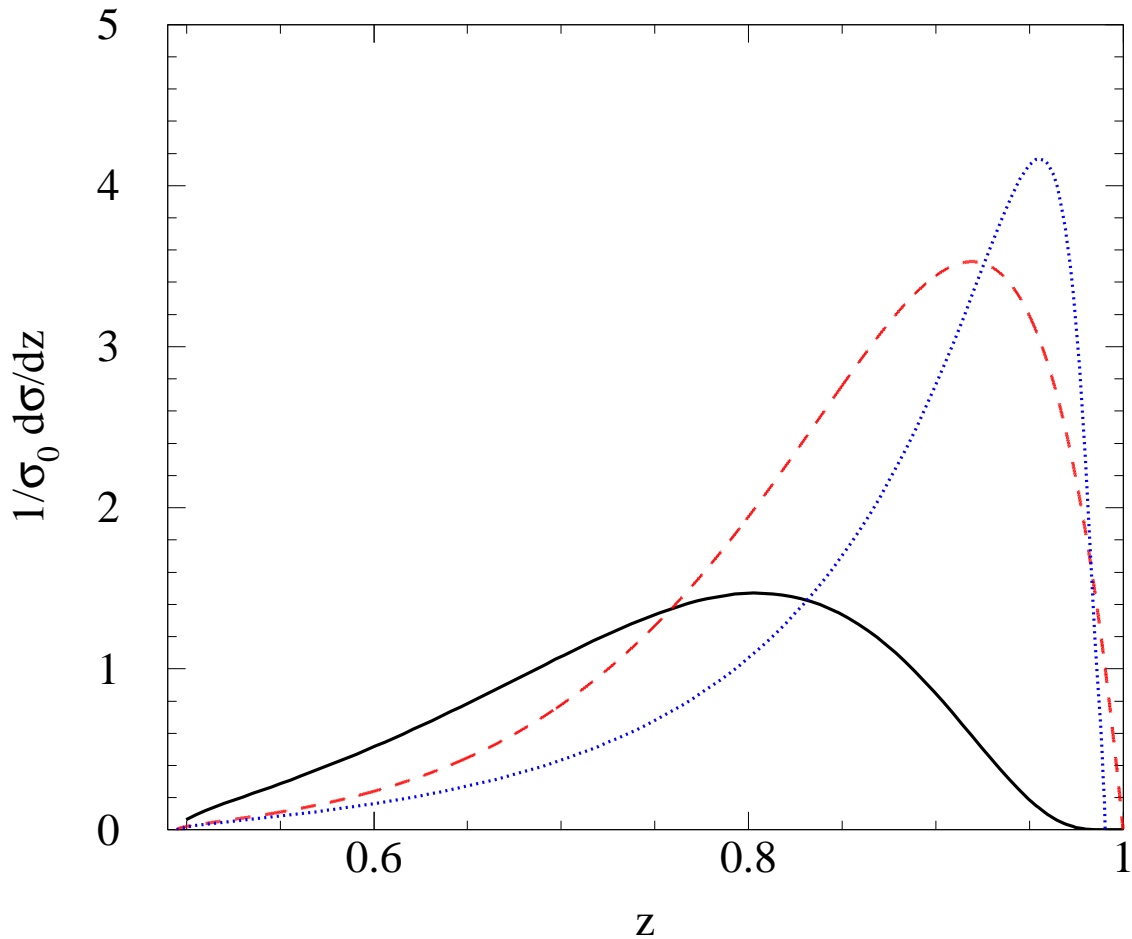}
\end{center}
\caption{\it  The color-octet contribution at the endpoint normalized to $\sigma_0$. The solid line is the perturbative resummation convoluted with the shape function. 
The dotted line is perturbative resummation only, and the dashed line is no resummation, but including a shape function.} 
\label{resumeff}
\end{figure}

In Fig.~\ref{resumeff} we show as the solid line the shape function in Eq.~(\ref{sfmodel}) convoluted with the perturbative resummed result, normalized to
$\sigma_0$.
The dashed line is a plot of the shape function alone, and the dotted line includes only the perturbative resummation. Here we use $m_c = 1.4$ GeV,   the one
loop coupling constant is evaluated at the hard scale $\mu_H = (s/M)(1-r)$, with $\Lambda_{\rm QCD} = 190 \, {\rm MeV}$, and  $a=1$ and $b = 2$. The value of the
first and second moments of the shape function for this choice of parameters are $890 {\rm\ MeV}$ and $(985 {\rm\ MeV})^2$ respectively.  We have chosen
the shape function parameters to give a reasonable description of the $p_\psi$ distributions measured by BaBar and Belle. Since $m_c v^2 \approx 500$ MeV
the moments are consistent with the velocity scaling rules. 
Fig.~\ref{resumeff} gives a picture of the effects of both the perturbative resummation and the shape function. The perturbative resummation gives a result that is
highly peaked in the endpoint region. The shape function is also peaked close to the endpoint, though at a lower value, and is broader. It is interesting to note
that the convoluted result is broader yet, and its peak is shifted to the left of both the perturbative resummed peak and the shape function peak. 

\begin{figure}
\begin{center}
\includegraphics[width=6.25in]{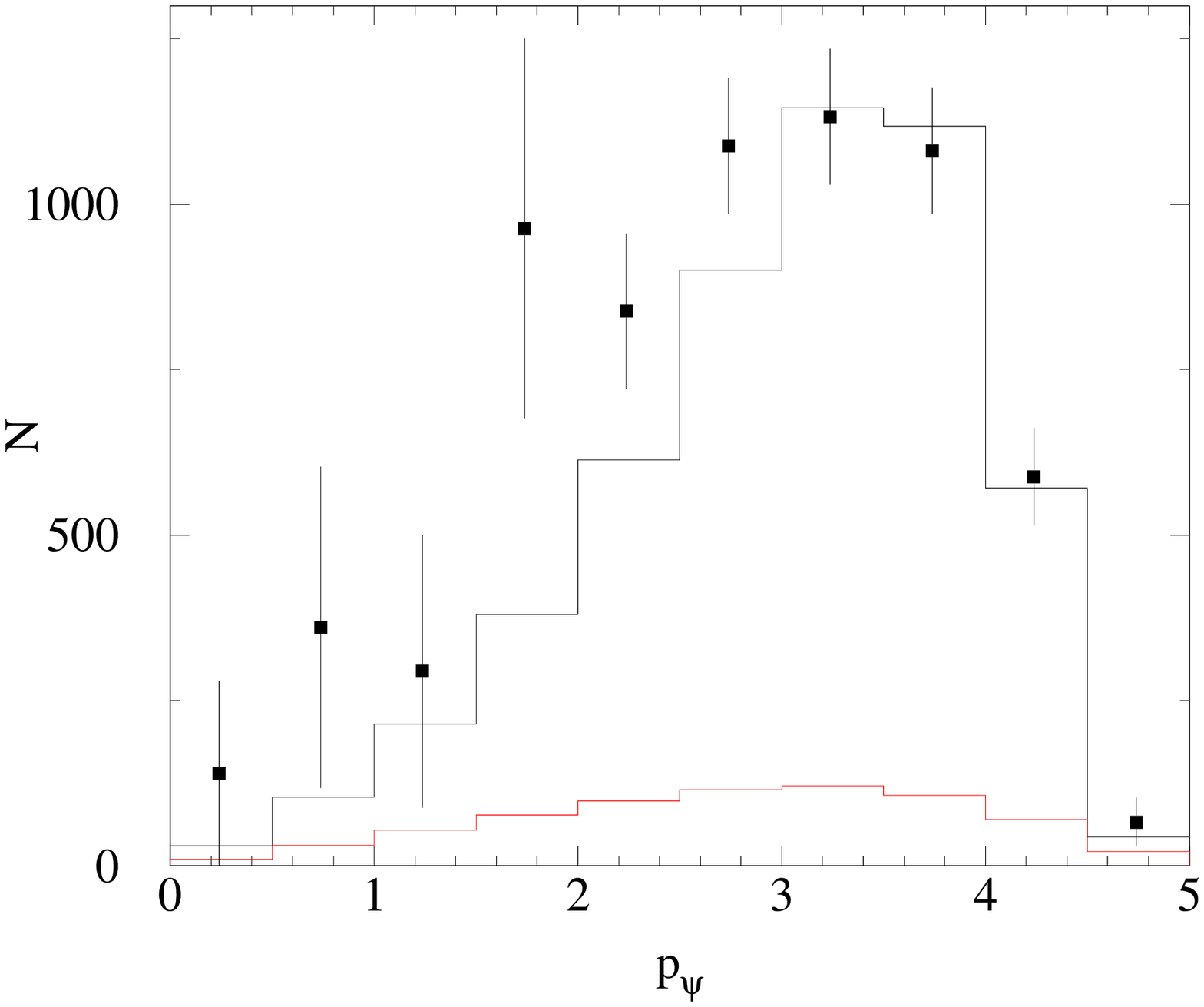}
\end{center}
\caption{\it  The sum of the color-octet and color-singlet contributions are plotted as the upper line. The lower line is the color-singlet contribution only, 
and the data are from the BaBar collaboration~\cite{Aubert:2001pd}. } 
\label{compbabar}
\end{figure}

\begin{figure}
\begin{center}
\includegraphics[width=6.25in]{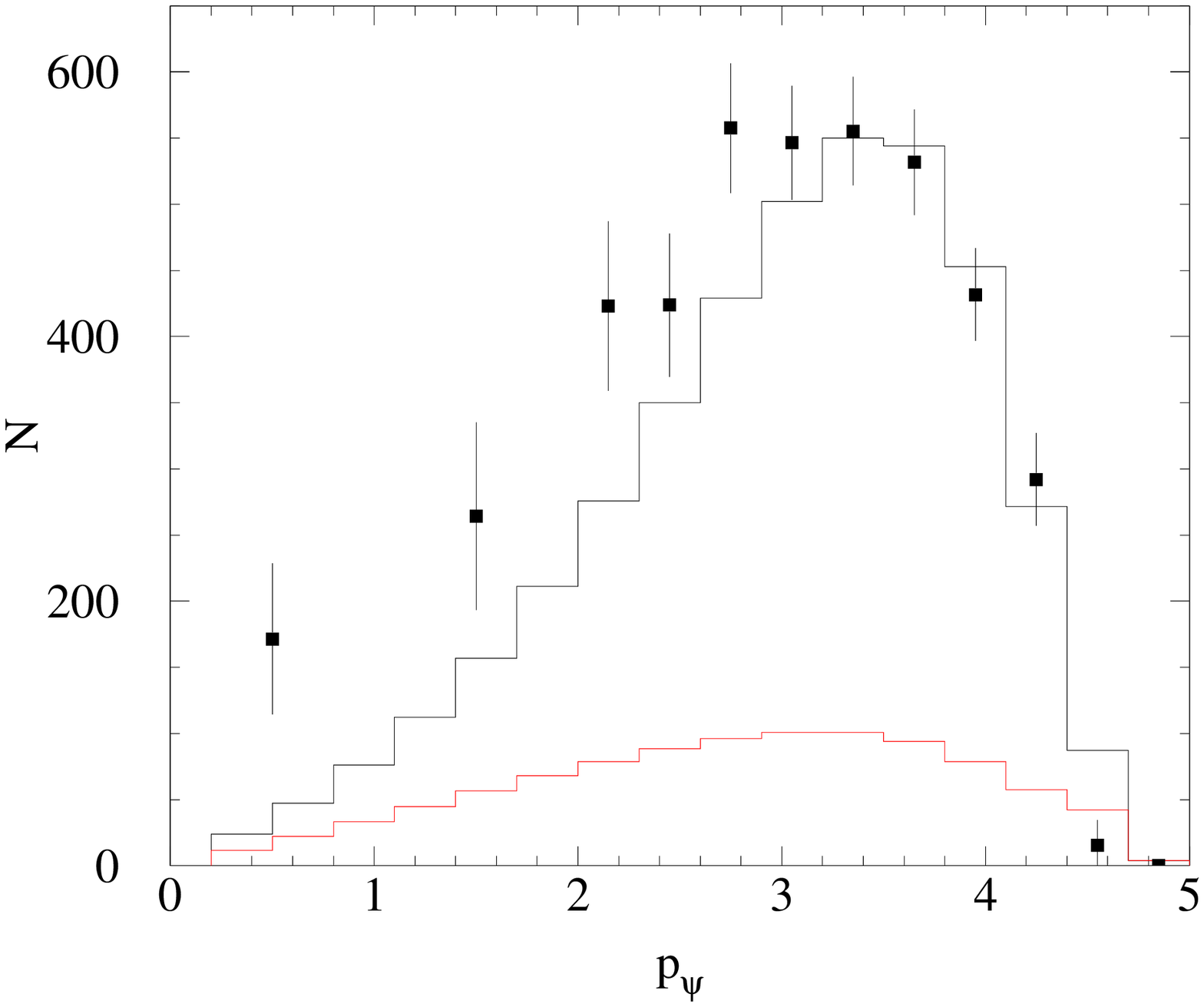}
\end{center}
\caption{\it  The sum of the color-octet and color-singlet contributions are plotted as the upper line. The lower line is the color-singlet contribution only, and 
the data are from the Belle collaboration~\cite{Abe:2001za}. } \label{compbelle}
\end{figure}

In order to make a consistent comparison of theory to data one needs to treat the endpoint of the color-singlet contribution in SCET and NRQCD. We leave this for
future work, and will use the leading order NRQCD calculation of the color-singlet contribution over the full range of momenta.  In Fig.~\ref{compbabar} we show the
sum of the color-octet and color-singlet contributions as the upper line, and the color-singlet contribution only as the lower line. The color-octet matrix elements
which set the normalization are chosen to be $\langle {\cal O}^\psi_8 (^1S_0)\rangle = \langle {\cal O}^\psi_8 (^3P_0)\rangle/m_c^2 = 1.3 \times 10^{-1}{\rm\
GeV}^3$. This is plotted against the BaBar data~\cite{Aubert:2001pd}.
In Fig.~\ref{compbelle} we show the same comparison to the Belle data~\cite{Abe:2001za}. The shape function parameters are the same, only the color-octet matrix
elements are smaller: $\langle {\cal O}^\psi_8 (^1S_0)\rangle = \langle {\cal O}^\psi_8 (^3P_0)\rangle/m_c^2 = 6.6 \times 10^{-2}{\rm\ GeV}^3$.  These values
of the color-octet matrix elements are consistent with data from photo- and hadroproduction~\cite{mefit}.
Figs.~\ref{compbabar} and \ref{compbelle} concisely present the central point of this work: when perturbative logarithms are resummed and the non-perturbative shape 
function is included the color-octet contribution becomes a broad distribution as a function of the $J/\psi$ momentum. Furthermore, for a reasonable choice of parameters
including the color-octet contribution gives a fairly good description of the $p_\psi$ distribution. The color-singlet contribution alone cannot describe the data on $J/\psi$ production at BaBar and Belle.

\begin{figure}
\begin{center}
\includegraphics[width=6.25in]{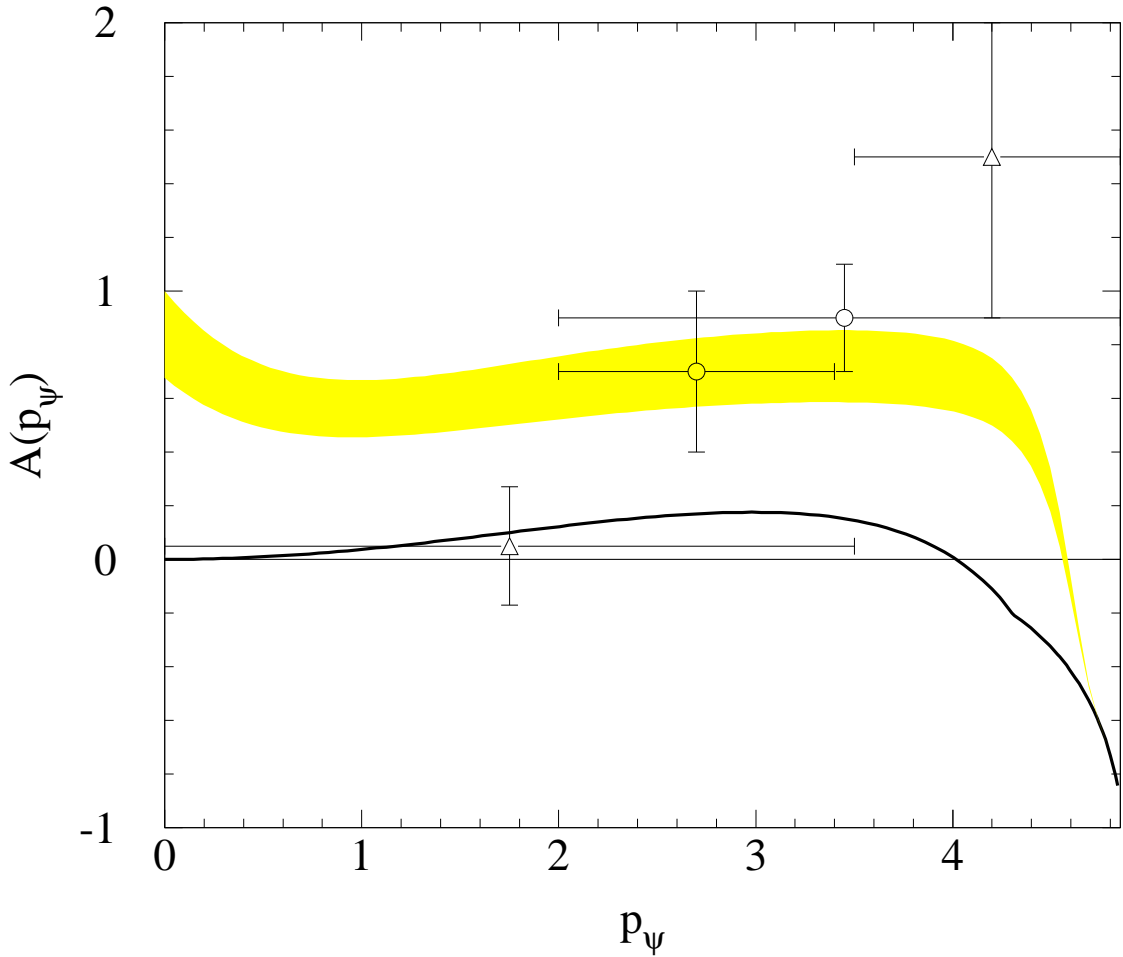}
\end{center}
\caption{\it  The angular distribution variable $A(p_\psi)$. The circles represent data from Belle~\cite{Abe:2001za} and the triangle data from BaBar~\cite{Aubert:2001pd}.
The solid line is the color-singlet contirbution alone while the shaded band includes the resummed color-octet contribution.} 
\label{aplot}
\end{figure}

The effect of the resummation on the angular distibution of the $\jpsi$ is shown in Fig.~\ref{aplot} where
we plot $A(p_\psi)$. The circles with error bars are data from Belle~\cite{Abe:2001za}
while the triangles with error bars are data from BaBar~\cite{Aubert:2001pd}. Note that the central value of one data
point lies outside the physical range, $-1 \leq A(p_\psi) \leq 1$. The solid line is the color-singlet prediction
and the shaded band includes the resummed color-octet contribution.  The color-octet $^1S_0$ production mechanism gives $A=1$
while the $^3P_J$ mechanisms gives $A\approx 0.7$. The band shown in Fig.~\ref{aplot} is obtained by 
varying the two matrix elements $\langle {\cal O}^\psi_8 (^1S_0)\rangle$
and $\langle {\cal O}^\psi_8 (^3P_0)\rangle$ subject to the constraint that both matrix elements are positive
and that 
\bea
\langle {\cal O}^\psi_8 (^1S_0)\rangle + 3.8 \langle {\cal O}^\psi_8 (^3P_0)\rangle/m_c^2 = 3.2 \times 10^{-2}{\rm\ GeV}^3,
\eea
which is the normalization of the curve shown in Fig.~\ref{compbelle}. If the normalization in Fig.~\ref{compbabar}
is used instead, the prediction for $A(p_\psi)$ is basically the same. The resummed color-octet contributions give $A(p_\psi) \approx 1$
for nearly the entire $p_\psi$ range. This is because for our choice shape function parameters the color-octet dominates
color-singlet over the entire range of $p$. The resummed color-octet calculation agrees with the Belle data as well
as the BaBar measurement for $p_\psi > 3.5$ GeV. However, the BaBar measurement for $p_\psi < 3.5$ GeV is about 2$\sigma$
below theory. 

Note that the factorization theorem we derive in this paper is only valid in the endpoint region 
and the predictions in the region $p_\psi < 2$ GeV should not be taken too seriously. The effects that are being 
resummed are no longer dominant in this region and other color-octet production mechanisms may be important. On the 
other hand, the small value of $A(p_\psi)$ could indicate the color-octet shape function we are using is too broad and some other mechanism
accounts for the observed $\jpsi$ production at these lower values of $p_\psi$.  
Though the results presented here give a fair description of the $p_\psi$ distribution, the Belle collaboration finds that roughly 60\% of the cross
section comes from the production of a $J/\psi$ in association with another $c \bar{c}$ pair. The theoretical prediction is that this contribution is less than 10\% of the cross
section. Clearly we do not resolve this issue here, and the mechanism responsible for the copious production of $J/\psi + D \bar{D}$ must be understood before any
measurements of the octet shape function parameters can be made at BaBar or Belle.

\section{Conclusion} 

Leading order NRQCD calculations of the $\jpsi$ spectrum in NRQCD predict an enhancement at maximal energy due to color-octet contributions.  Furthermore, the
angular dependence of the $\jpsi$ production due to color-octet contributions is markedly different than the color-singlet.   Recent experimental
investigations~\cite{Abe:2001za,Aubert:2001pd} of $e^+e^- \to \jpsi+X$ are at odds with the leading order predictions.  Large Sudakov logarithms appear in the higher
order corrections to the color-octet production mechanism which invalidate the perturbative expansion.  In addition, as  pointed out by Beneke, Rothstein, and
Wise \cite{Beneke:1997qw}, the nonperturbative $v^2$ expansion of NRQCD also breaks down at the endpoint, which necessitates the introduction of nonperturbative
shape functions to resum these large corrections.

We have studied the color-octet contribution to $e^+e^- \to \jpsi+X$ using SCET coupled to NRQCD.  First we showed how the differential rate factorizes
into a hard coefficient, a jet function containing collinear degrees of freedom, and a shape function.  The renormalization group equations of SCET were
used to move all large logarithms out of operators and into coefficient functions. This allowed the resummation of leading and next-to-leading Sudakov
logarithms, and naturally introduced the shape function into the differential spectrum.  While previous calculations of Sudakov logarithms in SCET have
employed a two-step evolution, we did a one-step evolution, similar to the velocity renormalization group in NRQCD~\cite{Luke:2000kz}.  Due to the
factorization of collinear and usoft degrees of freedom~\cite{Bauer:2001yt}, these two methods give identical results at leading order in the SCET power
counting.

Unfortunately, the shape functions for $\jpsi$ production are not known, so our calculation is not predictive.  In the future it may be possible to
extract the shape functions from either this spectrum, or from photo- or electroproduction, to be used as input in an analysis of one of the other
processes.  Until that time, the best we can do is to make a phenomenological model.  We are guided in this effort by knowledge of
the scalings of the moments of the shape function, which are given by NRQCD power counting rules.  

Using a simple model, we can obtain a qualitative feeling for the effects of the perturbative and nonperturbative resummations.  We found that both the
perturbative resummation or the inclusion of a shape function cause the differential spectrum to be lowered and shifted to smaller energies.  Together,
the spectrum is very broad, extending the color-octet contribution to very low energies.  Qualitatively, this could improve agreement with experimental data 
for the total rate, the $p_\psi$ distribution, and the angular dependence of the spectrum.  However, there are still many puzzles with the $\jpsi$ data, in particular the
large cross section for $\jpsi$  produced along with open charm and the exclusive double charmonium production~\cite{Abe:2002rb}, both of which
exceed existing theoretical estimates. In addition, there are discrepancies between the existing measurements by Belle and BaBar regarding the overall
normalization of the cross section as well as the polarization of the $J/\psi$ at largest values of $p_\psi$. These discrepancies must be resolved
and an understanding of the production mehanism which gives rise to the large  $\jpsi c \bar c$ 
cross section is needed before color-octet shape functions can be reliably extracted from the data.

\acknowledgments 
We would like to thank  Christian Bauer, Iain Stewart and Bruce Yabsley for helpful discussions.  S.F.~was supported in part by the Department of Energy under grant
number DOE-ER-40682-143.  A.L.~was supported in part by the Department of Energy under grant number DE-AC02-76CH03000.  T.M.~was supported in part by the Department
of Energy under grant numbers DE-FG02-96ER40945 and DE-AC05-84ER40150.

\appendix
\section{Deriving Factorization Theorem for the Endpoint}\label{factproof}

In this Appendix we show that the rate $e^+e^-\to J/\psi +X$ in the endpoint region can be factored into a hard coefficient, a
jet function, and a shape function. The jet and shape function involve only collinear and usoft fields, respectively.
The analysis is similar to the derivation of factorization 
for $b\to X_s \gamma$~\cite{Bauer:2001yt,Bauer:2002nz}.  We begin by writing the differential cross section as
\begin{eqnarray}
2 E_\psi \frac{d\sigma}{d^3p_\psi} &=& \frac{e^2}{16\pi^3 s^3}
L^{\mu\nu} 
\sum_X \langle 0 | J^\dagger_\nu(0) | J/\psi+X\rangle \langle J/\psi+X | J_\mu(0) | 0\rangle
(2\pi)^4 \delta^4(q-p_\psi-p_X)\nonumber\\
&=& \frac{e^2}{16\pi^3 s^3}
L^{\mu\nu} 
\int d^4 y\, e^{-i q\cdot y}
\sum_X \langle 0 | J^\dagger_\nu(y) | J/\psi+X\rangle 
\langle J/\psi+X | J_\mu(0) | 0\rangle \nonumber \\
&\equiv&\frac{e^2}{16\pi^3 s^3} L^{\mu\nu} T_{\mu\nu},
\end{eqnarray}
where the sum includes integration over the phase space of $X$.  The lepton tensor is
\begin{equation}
L^{\mu\nu} = p_1^\mu p_2^\nu + p_1^\nu p_2^\mu - g^{\mu\nu} p_1\cdot p_2,
\end{equation}
where $p_{1,2}$ are the momenta of the electron and positron, respectively.

Matching the QCD current $J_\mu$ onto SCET operators to leading order in $\lambda$ is discussed in Section \ref{fact}.
Note that $J^\dagger_\nu(y)$ picks up an additional phase due to the field redefinition relating QCD fields
and SCET/NRQCD effective theory fields:
\begin{equation}\label{effcurrents}
J^\dagger_\nu(y) = e^{i (M v - \bnP \, n/2)\cdot y} 
\left[
\psi^\dagger_{\bf p}\, \Gamma^{\dagger (8,{}^1S_0)}_{\beta \nu}\,  B^\beta_\perp \, \chi_{-{\bf p}}
+ \psi^\dagger_{\bf p}\, \Gamma^{\dagger (8,{}^3P_J)}_{\beta \nu \sigma \delta}  B^\beta_\perp
\Lambda\cdot\widehat{{\bf p}}^\sigma \Lambda\cdot\bsigma^\delta \chi_{-{\bf p}}
\right].
\end{equation}
The currents in Eq.~(\ref{effcurrents}) are the same that appear in inclusive radiative $\Upsilon$ decay
\cite{Bauer:2001rh}.  To lowest order in $\alpha_s$, $\Gamma^{\dagger (8,{}^1S_0)}_{\beta \nu}$ is the same as in radiative 
$\Upsilon$ decay, while $\Gamma^{\dagger (8,{}^3P_J)}_{\beta \nu \sigma \delta}$ is different.

We will only derive the factorization theorem for the $^1S_0$ contribution. 
The generalization to the $^3P_J$ contribution is straightforward.  The
$^1S_0$ contribution to the hadronic tensor $T_{\mu\nu}$ can be written as
\begin{eqnarray}\label{tmunu}
T_{\mu\nu} &=& \int d^4 y\, e^{-i\sqrt{s}/2(1-\hat{x})\bn\cdot y} 
\Gamma^{(8,{}^1S_0)}_{\alpha \mu}\Gamma^{\dagger(8,{}^1S_0)}_{\beta \nu} \nonumber \\
&& \phantom{\int d^4 y\, e^{i\sqrt{s}/2(1-x)\bn\cdot y}}\times
\sum_X \langle 0 | \chi^\dagger_{-{\bf p}} 
B^\beta_\perp \psi_{\bf p}(y) | J/\psi+X\rangle 
\langle J/\psi+X | \psi^\dagger_{\bf p'}  
B^\alpha_\perp \chi_{-{\bf p'}}(0) | 0\rangle \nonumber\\
&\equiv& \Gamma^{(8,{}^1S_0)}_{\alpha \mu}\Gamma^{\dagger(8,{}^1S_0)}_{\beta \nu}
T^{\beta\alpha}_{\rm eff}.
\end{eqnarray}
In the exponent of Eq.~(\ref{tmunu}), we  have used $\bnP B_\perp^\alpha = - \sqrt{s}(1 - r) B_\perp^\alpha$ so
\begin{eqnarray}
q^\mu - M v^\mu + \bnP \frac{n^\mu}{2}  = \sqrt{s}(1 - \hat{x}) \frac{\nb^\mu}{2} - \frac{r \sqrt{s}}{\hat{x}}(1 - \hat{x}) \frac{n^\mu}{2} 
 \, .
\end{eqnarray}
The term proportional to $n^\mu$ is suppressed by $r \approx 0.08$ and can be neglected. 
We can now decouple the usoft gluons in $T_{\rm eff}^{\beta \alpha}$ using the field  redefinition \cite{Bauer:2001yt}
\begin{equation}\label{fieldredef}
A^\mu_{n,q} = Y A^{(0) \mu}_{n,q} Y^\dagger
\hspace{.5cm} \to \hspace{.5cm} 
W_n = Y W_n^{(0)} Y^\dagger \,,
\end{equation}
where the first identity implies the second.  The collinear fields with the superscript $(0)$ do not interact with usoft fields
to lowest order in $\lambda$. After this field redefinition the color-octet ${}^1S_0$ current becomes
\begin{equation}
\tilde{J}^\alpha_{(8,{}^1S_0)}  = \psi^\dagger_{\bf p} Y 
 B^{(0) \alpha}_\perp  Y^\dagger \chi_{-{\bf p}} \,.
\end{equation}
The $\jpsi$ does not contain any collinear
quanta,\footnote{If the $J/\psi$ is produced with large enough energy, the charm quarks could be considered collinear particles
though the Lagrangian is modified by the presence of the charm quark mass \cite{Leibovich:2003jd}.   However, at very large energies, for 
instance at LEP, quark fragmentation dominates over the contributions considered here \cite{LEP}.} so we can write 
\begin{eqnarray}\label{bigTeff}
T^{\beta\alpha}_{\rm eff} &=& \nonumber \\
& & \hspace{-4ex} \int d^4 y\, e^{-i\sqrt{s}/2(1-\hat{x})\bn\cdot y}
\langle 0 | \chi^\dagger_{-{\bf p}} Y T^A Y^\dagger \psi_{\bf p} (y) 
\sum_{X_u} | J/\psi+X_u \rangle \langle J/\psi+X_u | 
\psi^\dagger_{\bf p} Y T^A Y^\dagger \chi_{-{\bf p}}(0) | 0\rangle \nonumber\\
&&\phantom{\int d^4 y\, e^{i\sqrt{s}/2(1-\hat{x})\bn\cdot y}}
\times \frac12 \langle 0 | {\rm Tr} \left[ T^B B^{(0) \beta}_{\perp}(y) \right] 
\sum_{X_c} | X_c \rangle \langle X_c |
{\rm Tr}\left[ T^B B^{(0)\alpha}_{\perp}(0) \right]  | 0 \rangle \,,
\end{eqnarray}
where $X_u$ contains only usoft particles and $X_c$ contains collinear particles.   This is possible since interpolating fields for
all final state particles can be written with  either all collinear or all usoft quanta.   By introducing an interpolating field,
$a_\psi$, for the $\jpsi$,  we can use the completeness of states in the usoft and collinear sectors to write
\begin{eqnarray}\label{ucomplete}
\sum_{X_u} | J/\psi+X_u \rangle \langle J/\psi+X_u | &=& 
a_\psi^\dagger \sum_{X_u} | X_u \rangle \langle X_u | a_\psi = a_\psi^\dagger a_\psi, \\
\label{ccomplete}
\sum_{X_c} | X_c \rangle \langle X_c | &=& 1.
\end{eqnarray}
Using Eqs.~(\ref{ucomplete}) and (\ref{ccomplete}) in Eq.~(\ref{bigTeff}), we get
\begin{eqnarray}\label{tba}
T^{\beta\alpha}_{\rm eff} &=&
\frac12 \int d^4 y\, e^{-i\sqrt{s}/2(1-\hat{x})\bn\cdot y}
\langle 0 | \chi^\dagger_{-{\bf p}} Y T^A Y^\dagger \psi_{\bf p} (y) 
\,a_\psi^\dagger a_\psi\,
\psi^\dagger_{\bf p} Y T^A Y^\dagger \chi_{-{\bf p}}(0) | 0\rangle \nonumber\\
&&\phantom{\frac12 \int d^4 y\, e^{i\sqrt{s}/2(1-x)\bn\cdot y}}
\times \langle 0 | {\rm Tr} \left[ T^B B^{(0) \beta}_{\perp}(y) \right] 
{\rm Tr}\left[ T^B B^{(0)\alpha}_{\perp}(0) \right]  | 0 \rangle \,.
\end{eqnarray}

Next it is useful to define the jet and shape functions. The jet function is defined by
\begin{equation}\label{jetdef}
\langle 0 | {\rm Tr} \left[ T^B B^{(0) \beta}_{\perp}(y) \right] 
{\rm Tr}\left[ T^B B^{(0)\alpha}_{\perp}(0) \right]  | 0 \rangle
\equiv -\frac{4\pi g_\perp^{\alpha\beta}}{s(1+r)} \int \frac{d^4k}{(2\pi)^4} e^{-i k\cdot y} 
J(\bn\cdot p \,n\cdot k + p_\perp^2).
\end{equation}
Note that the jet function is a function of the labels $\nb\cdot p$ and $p_\perp$ of the collinear fields in $B_\perp^{(0)\alpha}$,
and the residual momentum $n\cdot k$. For the process we are interested in $\nb\cdot p = -\nb \cdot p_X$ and $p_\perp=0$.
Since the jet function is independent of $k_\perp$ and $\nb \cdot k$, the momentum integration
over these components yields $\delta(n\cdot y)$ and $\delta^2(y_\perp)$. The identity (valid for $q_0 > 0$)
\begin{eqnarray}
\int d^4 y \, e^{i q\cdot y} \, \langle 0 \!\!\!\!&|& \!\!\!\!{\rm Tr} \left[ T^B B^{(0) \beta}_{\perp}(y) \right] 
{\rm Tr}\left[ T^B B^{(0)\alpha}_{\perp}(0) \right]  | 0 \rangle \nonumber \\
&=& 2 \,{\rm Im} \left[i \int d^4y \, e^{i q\cdot y} \,
\langle 0 | T \left( {\rm Tr} \left[ T^B B^{(0) \beta}_{\perp}(y) \right] 
{\rm Tr}\left[ T^B B^{(0)\alpha}_{\perp}(0) \right]  \right) | 0 \rangle  \right] ,
\end{eqnarray}
shows that the jet function is related to the imaginary part of $T-$ordered
products of the composite field $B_\perp^{(0)\alpha}$, which is useful for explicit computations.

The shape function is 
\begin{eqnarray}\label{shapedef}
S^{(8,^1S_0)}(\ell^+) &=& \int \frac{dy^-}{4\pi} e^{-\frac{i}{2} \ell^+ y^-}
\frac{
\langle 0 | \chi^\dagger_{-{\bf p}}Y T^B Y^{\dagger} \psi_{\bf p} (y^-)  
\,a_\psi^\dagger a_\psi\, \psi^\dagger_{\bf p}Y T^B Y^{\dagger}\chi_{-{\bf p}}(0) | 0\rangle
}{4 m_c\langle {\cal O}^\psi_8(^1S_0)\rangle} \\
&=& \frac{
\langle 0 | \chi^\dagger_{-{\bf p}} T^B \psi_{\bf p} 
\,a_\psi^\dagger a_\psi\, \delta(\ell^+ - i n\cdot D)
\psi^\dagger_{\bf p} T^B \chi_{-{\bf p}} | 0\rangle
}{4 m_c\langle {\cal O}^\psi_8(^1S_0)\rangle} \, . \nonumber
\end{eqnarray}
The second line of Eq.~(\ref{shapedef}) is most easily derived by evaluating the first expression in $n\cdot A_{us}=0$
gauge (where $Y=Y^\dagger =1$), then replacing ordinary derivatives by usoft gauge covariant derivatives to restore
usoft gauge invariance. The normalization is defined so that 
\begin{eqnarray}
\int d\ell^+ S^{(8,^1S_0)}(\ell^+) =
\frac{\langle 0 | \chi^\dagger_{-{\bf p}} T^B \psi_{\bf p} 
\,a_\psi^\dagger a_\psi\, \psi^\dagger_{\bf p} T^B \chi_{-{\bf p}} | 0\rangle
}{4 m_c\langle {\cal O}^\psi_8(^1S_0)\rangle} = 1.
\end{eqnarray}
This function depends on the state of the $c\bar{c}$ pair in the $\jpsi$, and thus there will be a different shape function for the
$^3P_J^{(8)}$ contribution. 

We substitute these expressions into Eq.~(\ref{tba}), then use
\begin{eqnarray}
\int \frac{d^3p_\psi}{2 E_\psi} (1+\cos^2\theta)
= \frac{2\pi}{3}{s(1-r^2)} P[r,z]dz  \, ,
\end{eqnarray}
where $\cos \theta$ is the angle of the $J/\psi$ with the $e^+e^-$ beam
and $P[r,z] = \sqrt{(1+r)^2 z^2 -4 r}/(1-r) $. Note that $P[r,1]=1$. In the phase space we use
the short distance mass $2 m_c$ not $M_\psi$. The result for the differential cross section is
\begin{eqnarray}
\frac{d\sigma^{(8,^1S_0)}}{dz} &=& \sigma_0^{(8,^1S_0)} P[r,z]
\int d\ell^+ S^{(8,^1S_0)}(\ell^+) J \left(
\sqrt{s}(1 -r) [\sqrt{s} (1- x + \bar \Lambda/M )- \ell^+  ] \right)\nonumber \\
&=& \sigma_0^{(8,^1S_0)} P[r,z]
\int d\hat\ell^+ S^{(8,^1S_0)}(\hat\ell^+) J \left[
s(1 -r) (1- x + \bar \Lambda/M - \hat\ell^+/M  ) \right].
\end{eqnarray}
Here we have used the fact that $d\ell^+ S^{(8,^1S_0)}(\ell^+)$ is boost invariant, and
\begin{eqnarray}
p^\mu_X = \frac{\sqrt{s}}{2} \Bigg[ \Bigg(1 - \frac{r}{\hat{x}} \Bigg) n^\mu  
+ (1- \hat{x}) \bar{n}^\mu \Bigg]  - \ell^\mu \, ,
\end{eqnarray}
so
\begin{eqnarray}
p_X^2 &=& \sqrt{s} \Bigg( 1 - \frac{r}{\hat{x}} \Bigg) \left(
\sqrt{s}(1- \hat{x}) - \ell^+\right)  \nonumber \\
 &=& s(1 -r )
\left(1- x +\frac{\bar \Lambda}{M} -\frac{\hat\ell^+}{M}  \right)   \, ,
\end{eqnarray}
where $\ell^+ =\hat\ell^+x\sqrt{s}/M_\psi$. 
In the last line we have expanded $p_X^2$ to lowest order in $1-x$ and $\bar \Lambda/M$. Finally, writing 
$z$ in terms of $x$ 
\begin{eqnarray}
z = \frac{s x + M_\psi^2/x}{s+M_\psi^2} \approx 1 -\frac{1-r}{1+r}(1-x),
\end{eqnarray}
where again we have expanded to lowest order in $1-x$ and $\bar \Lambda/M$, and defining 
\begin{eqnarray}\label{lhatxi}
\hat\ell^+ = \bar\Lambda +  M\frac{1+r}{1-r}(1-\xi) ,
\end{eqnarray}
we find $p_X^2= s(1+r)(\xi - z)$ . The factorization
theorem can then be written as 
\begin{eqnarray}
\frac{d\sigma^{(8,^1S_0)}}{dz} &=& \sigma_0^{(8,^1S_0)} P[r,z]\, M\frac{1+r}{1-r}
\int_z^1 d\xi\, S^{(8,^1S_0)}\left[M\frac{1+r}{1-r} (1-\xi)+\bar\Lambda\right] J (s(1+r)(\xi-z))  
\nonumber\\
&\equiv& \sigma_0^{(8,^1S_0)} P[r,z] 
\int_z^1 d\xi\, S^{(8,^1S_0)}(\xi) J (s(1+r)(\xi-z)),
\end{eqnarray}
where $S^{(8,^1S_0)}$ as a function of $\xi$ 
is defined such that $\int d\xi {S}^{(8,^1S_0)}(\xi) = 1$.
The integration limits on $\xi$ are easy to understand: $p_X^2 > 0$ requires $\xi > z$
and the upper limit comes from the requirement that $d\sigma/dz$ vanish at the kinematic limit $z=1$.

\section{The Jet Function}\label{jetapp}
In this section we discuss the jet function and its renormalization.
We show how to renormalize the moments of the jet function and evaluate
these moments to $O(\alpha_s)$. We derive the renormalization group equation (RGE) for the moments needed for the resummation
of endpoint logarithms.  We will assume that the usoft fields have already been decoupled by the field redefinition, Eq.~(\ref{fieldredef}), and will use the superscript $(0)$ to denote bare fields instead.

We can invert Eq.~(\ref{jetdef}) to obtain
\begin{eqnarray}\label{topjetdef}
J(\nb \cdot p \, n\cdot k +p_\perp^2) 
&=&  - \frac{s(1+r)}{4 \pi}\\
&&\times{\rm Im} \left[i \int d^4y \, e^{i k \cdot y} \,
\langle 0 | T \left( {\rm Tr} \left[ T^B B^{\beta}_{\perp}(y) \right] 
{\rm Tr}\left[ T^B B_{\perp\,\beta}(0) \right]  \right) | 0 \rangle  \right] .\nonumber
\end{eqnarray}
To lowest order in $\alpha_s$, we replace the $B_\perp^{\alpha}$ with $A_\perp^\alpha$, so up to a numerical factor
the jet function is given by the discontinuity of the SCET collinear gluon propagator. Therefore, 
\begin{eqnarray}\label{treejet}
J(\nb \cdot p \, n\cdot k +p_\perp^2) 
&=&   -\frac{s(1+r)}{ \pi}\,{\rm Im} \frac{1}{\nb \cdot p \, n \cdot k + p_\perp^2 + i \epsilon} \\
&=&  s(1+r)\delta(\nb \cdot p\,  n \cdot k + p_\perp^2 ) \nonumber \\
&=&  \delta(\xi - z) \nonumber \, .
\end{eqnarray}
In the last line we have used $p_X^2 = \nb \cdot p \, n \cdot k + p_\perp^2 =s(1+r)(\xi -z)$.

\begin{figure}
\begin{center}
\picspacehalf
\includegraphics[width=6in]{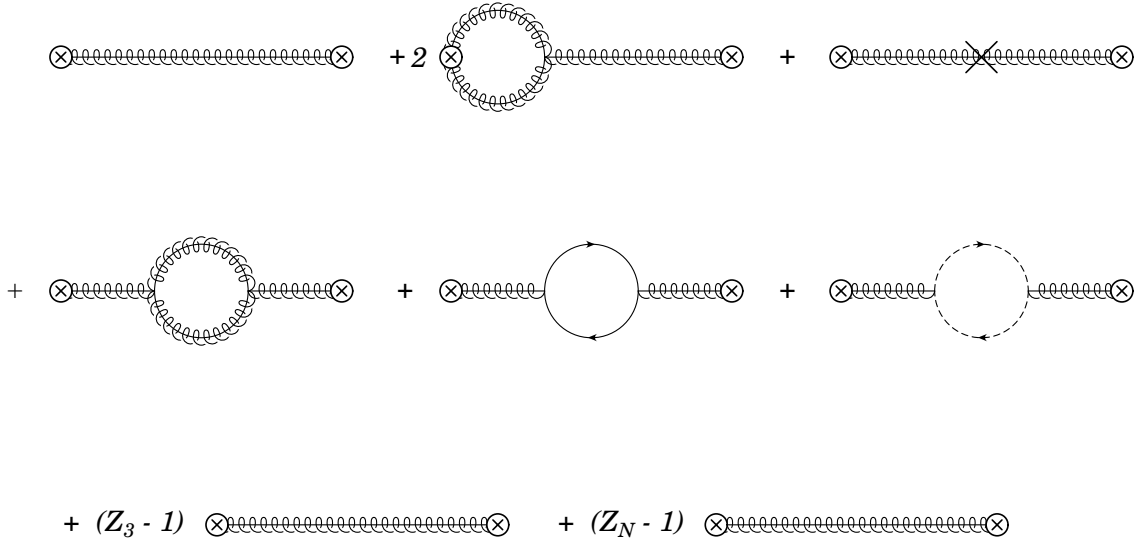}
\picspace
\caption{\it Graphs contributing to $J_N$ up to $O(\alpha_s)$. 
\label{jetfig}}
\end{center}
\end{figure}

To calculate the renormalization group equations for the moments of the jet function, we begin with the definition  in Eq.~(\ref{topjetdef}). The
moments as defined in Eq.~(\ref{moments}) do not correspond to local SCET operators, so it is easiest to compute the
time-ordered product to one-loop then  take moments of the result. Since the divergence depends on the moment
variable, $N$, we need to introduce an $N$ dependent renormalization for the moments of the jet function: $J_N  =
Z_N J_N^{(0)}$, where $J_N^{(0)}$ is the $N$th moment of the bare jet function. The field appearing in the definition of
$J_N$ is the composite  field $B_\perp$. For computing the $O(\alpha_s)$ anomalous dimension one can simply
ignore  the $Z$ factors for the fields in the Wilson lines since these counterterms will only contribute at
$O(\alpha_s^2)$. Therefore, $B_\perp = \sqrt{Z_3} B^{(0)}_\perp$, and
\begin{eqnarray}\label{renj}
J_N &=& - \frac{s(1+r)}{4 \pi} Z_N Z_3  \\
&&\times\int_0^1 dz\,z^N\,{\rm Im} \left[i \int d^4y \, e^{i k \cdot y} \,
\langle 0 | T \left( {\rm Tr} \left[ T^B B^{(0) \beta}_{\perp}(y) \right] 
{\rm Tr}\left[ T^B B^{(0)}_{\perp\,\beta}(0) \right]  \right) | 0 \rangle  \right] ,\nonumber
\end{eqnarray} 
where $Z_3$ is the wavefunction renormalization of the gluon field,
\begin{equation}
Z_3 = 1 + \frac{\alpha_s}{4\pi}\frac{1}{\epsilon} \left(C_A \frac53 - n_f \frac23\right).
\end{equation}
The $O(\alpha_s)$ corrections to the
time-ordered product are shown in Fig.~\ref{jetfig}. In addition to the  one-loop graphs there are three
counterterm contributions. Two come from the  $Z$ factors in Eq.~(\ref{renj}) and correspond to the
last two diagrams  in Fig.~\ref{jetfig}. There is also  a tree level insertion of the $O(\alpha_s)$ 
wavefunction counterterm for the collinear gluons, shown as a cross in the third diagram of 
the first line of Fig.~\ref{jetfig}.  This contribution cancels against the term proportional to
$Z_3-1$ in the third line of Fig.~\ref{jetfig}.

Therefore, the divergences from the one-loop graphs must be canceled by the diagram proportonal to
$Z_N-1$ in the last line of Fig.~\ref{jetfig}. Evaluation of the graphs and
derivation of the renormalization group equation is straightforward.  We already calculated the tree-level diagram in Eq.~(\ref{treejet})
and the self energy corrections are standard.  The only remaining graph to
calculate is the second diagram in Fig.~\ref{jetfig}, which has a divergent contribution of

\begin{equation}
-\frac{s(1+r)}{\pi}\int_0^1 dz\,z^N\,\frac{\alpha_s C_A}{2\pi}
{\rm Im}\left\{\frac{1}{p_X^2 + i \delta} \left[\frac{2}{\epsilon^2} + 
\frac{1}{\epsilon}\left(1 + 2 \log\frac{\bar\mu^2}{(-1-i\delta)p_X^2}\right)\right]\right\}.
\end{equation}
Taking the imaginary part, using
\begin{equation}
{\rm Im}\left[\frac{1}{p_X^2 + i \delta}  \log\frac{\bar\mu^2}{(-1-i\delta)p_X^2}\right] = 
\frac{\pi}{s(1+r)}\left[\frac1{(1-z)_+} - \delta(1-z)\log\frac{\bar\mu^2}{s(1+r)}\right],
\end{equation}
we can easily integrate over $z$.  Setting the divergent part to zero gives
\begin{equation}
Z_N -1 = -\frac{\alpha_sC_A}{\pi\epsilon^2} -
\frac{\alpha_s}{2\pi\epsilon}\left(\frac{11C_A}{6} -\frac{n_f}{3} + 
   2 C_A \log\frac{\bar\mu^2N}{s(1+r)N_0}\right).
\end{equation}
Calculating the anomalous dimension in the usual way, we get Eq.~(\ref{jrge}) for the renormalization of the jet function.
We can also evaluate the one-loop expression for the renormalized jet function from the diagrams in Fig.~\ref{jetfig}:
\begin{equation}
J_N  = 1 + \frac{C_A \alpha_s}{2\pi}\log^2 \left(\frac{{\bar \mu}^2 N}{s(1+r)N_0}\right)
+\frac{\alpha_s}{\pi}\left(\frac{11}{12} C_A -\frac{1}{6}n_f\right) \log \left(\frac{{\bar \mu}^2 N}{s(1+r)N_0}\right)
+ ... \,
\end{equation}
The ellipsis represents contributions which are not enhanced by large logarithms. Note that the result of explicit evaluation of $J_N$ agrees with the solution of
the renormalization group equation to $O(\alpha_s)$.  It is clear that the large logs in the moments of the jet function are minimized when ${\bar \mu}^2 =
s(1+r)N_0/N$.


\end{document}